\documentclass[prb,showpacs,unsortedaddress]{revtex4}
\usepackage[T1]{fontenc} 
\usepackage[utf8]{inputenc}
\usepackage{appendix}
\usepackage{tgbonum}
\usepackage{amsmath,amssymb}

\usepackage{lmodern}
\usepackage[pdftex]{graphicx}
\usepackage[caption=false]{subfig}
\usepackage[capitalise]{cleveref}
\usepackage{lmodern}
\usepackage{bbm}
\usepackage{tikz}
\usepackage{circuitikz}
\usepackage{braket}
\usepackage[caption=false]{subfig}

\begin{document}
\title{AC driven strongly correlated quantum circuits and Hall edge states: Unified photo-assisted noise and revisited minimal excitations}

\author{In\`es Safi}

\address{Laboratoire de Physique des Solides (UMR 5802),\\ CNRS-Universit\'e
Paris-Sud and Paris-Saclay,
B\^atiment 510, 91405 Orsay, France}

\today
\begin{abstract}
We study the photo-assisted noise generated by time-dependent or random sources and transmission amplitudes. We show that it obeys a perturbative non-equilibrium fluctuation relation that fully extends the lateral-band transmission picture in terms of many-body correlated states. This relation holds in non-equilibrium strongly correlated systems such as the integer or fractional quantum Hall regime as well as in quantum circuits formed by a normal or Josephson junctions strongly coupled to an electromagnetic environment, with a possible temperature bias. We then show that the photo-assisted noise is universally super-poissonian, giving an alternative to a theorem by L. Levitov {\it et al} which states that an ac voltage increases the noise. Restricted to a linear dc current, we show that the latter does not apply to a non-linear superconducting junction. Then we characterize minimal excitations in non-linear conductors by ensuring a poissonian photo-assisted noise, and show that these can carry a non-trivial charge value in the fractional quantum Hall regime. We also propose methods for shot noise spectroscopy and for a robust determination of the fractional charge which is more advantageous than those we have proposed previously and  implemented experimentally.  \end{abstract}

 

\maketitle

\tableofcontents
\section{Introduction}
Time-dependent (TD) transport presents a powerful probe of quantum phenomena by introducing multiple parameters or functions under control: frequencies for emitted noise generated by constant forces, or time dependent forces generating current or noise at low or finite frequencies. \cite{glattli_photo,gabelli_08,gabelli_13,reulet_2017_Photo_DCB,dubois_minimization_integer,glattli_imen_2022} It has been analyzed in a mesoscopic context through seminal theoretical approaches, such as the Tien-Gordon theory \cite{tucker_rev,tomasz_09,thorwart_mix} or the Landauer-B\"uttiker scattering approach, associated with the Floquet theory. \cite{buttiker_traversal_time,blanter_buttiker,lesovik_photo,nazarov_book,vavenic_07,vavenic_13_photo,belzig_zhan_NJP_2020} The effect of a periodic ac voltage $V_{ac}(t)$ at a frequency $\Omega_0$ is often addressed within the so-called lateral-band transmission scheme, where $V_{ac}(t)$ is viewed as a coherent radiation with translates one electron energy by $l\Omega_0$ for each integer number $l$ of exchanged photons. This yields a relation between the induced low frequency shot noise and a superposition of duplicates over $l$ of the noise in the dc regime. By linking current fluctuations, we coin it as a fluctuation relation (FR), which we distinguish from fluctuation-dissipation relations that involve current or conductance. The noise induced by $V_{ac}$ is called photo-assisted shot noise (PASN); it should be higher than its value in the dc regime according to a theorem by L. Levitov {\it et al}.\cite{keeling_06_ivanov}  While poissonian shot noise in the dc regime is common to classical and quantum particles, the PASN has the interest to provide a signature of a quantum behavior through rectified current fluctuations. 

The PASN is also an important tool to explore remarkable collective phenomena and macroscopic manifestation of quantum physics when strong correlations play a crucial role, but for which the lateral-band transmission picture has been claimed to be inappropriate.\cite{photo_review} Though, such a picture was recovered within specific models. For instance, unexpected Tien-Gordon type relations were obeyed in the Tomonaga-Luttinger Liquid (TLL) model (a generic non-Fermi liquid arising in strongly correlated 1-D systems) either by the photo-assisted current (PAC), \cite{wen_photo_PRB_91,sassetti_99_photo} or by the PASN in Refs.[\onlinecite{crepieux_photo,martin_sassetti_prl_2017}] though not compared to its Tien-Gordon type form. Such works were indeed covered by our unifying NE perturbative approach. \cite{ines_cond_mat,ines_portier_2015,ines_eugene,ines_PRB_2019}

\begin{figure}[htbp]
\begin{center}
\includegraphics[width=6cm]{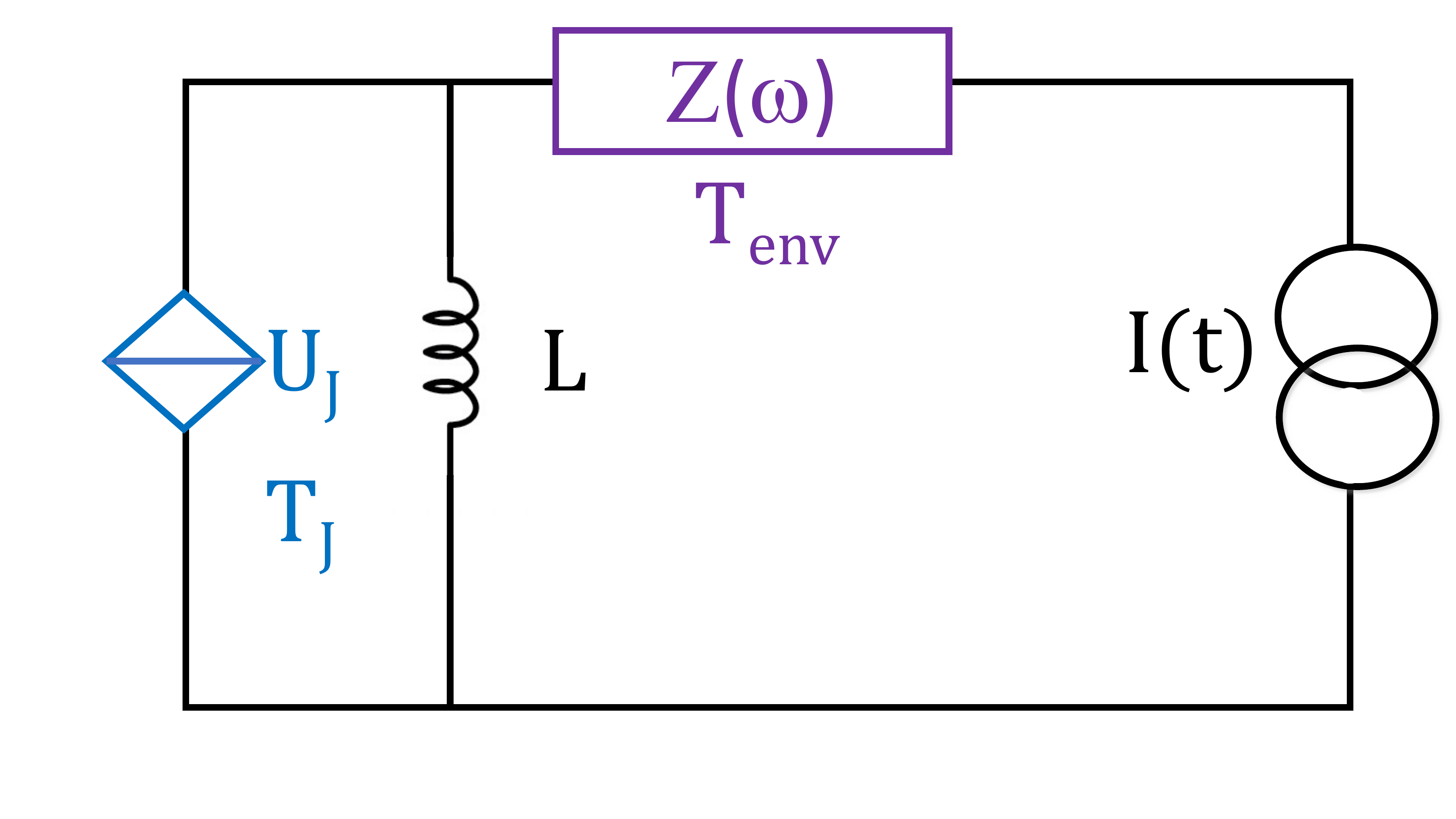}
\caption{\small Fourth example: a dual-phase Josephson junction with a small effective parameter $U_J$. Voltage and current are permuted, so that one imposes a time dependent current, while the voltage noise across the junction obeys the NE FR. Average voltage was computed in Ref.\onlinecite{photo_josephson_hekking} and found to obey the relation provided by the perturbative approach \cite{ines_eugene,ines_PRB_2019}.}
\label{fig_JJ_dual}
\end{center}
\end{figure}

Here the same approach is adopted to extend fully the lateral-band transmission picture for PASN to many-body correlated states. Contrary to a majority of studies restricted to periodic voltages, we also extend it to non-periodic tunneling amplitudes and voltages, which can then be generated by fluctuating sources or pseudo-random lorentzian pulses. \cite{glattli_pseudorandom_PRB_2018,degiovanni_PRXQuantum_2021} Thus the approach cannot be coined as Tien-Gordon theory. It unifies many previous works based on specific models, \cite{dubois_minimization_integer,crepieux_photo,martin_sassetti_prl_2017} beyond which it extends to a larger universality class of strongly correlated circuits and situations. Let's mention a quantum point contact (QPC) in compressible edge states in the integer quantum Hall effect (IQHE) or the fractional quantum Hall effect (FQHE) (see Fig.\ref{fig_Hall}), or a quantum circuit formed by a QPC, a Josephson junction (JJ) \cite{worsham_JJ_AC} (Fig.\ref{fig_JJ}) or a dual phase-slip JJ (Fig.\ref{fig_JJ_dual}) strongly coupled to an electromagnetic environment. 
Another strength of the NE approach is that it goes beyond initial thermalized many-body states to NE ones. It covers for instance the SIN junction in a NE diffusive wire studied in Ref.[\onlinecite{hekking_NE_SIN_PRB_2004}], or a quantum circuit with a temperature bias we've studied recently. \cite{ines_pierre_thermal_2021} In addition, this approach led to some fluctuation NE relations not derived so far, \cite{ines_PRB_R_noise_2020} even for independent electrons. Though we consider here the PASN of a current operator, the latter could refer, depending on the model, to a generalized force such as a voltage operator in the dual SIS junction \cite{photo_josephson_hekking} or a spin current in a magnetic junction.
 
The PASN is especially relevant for two rapidly growing and fascinating domains where injection and manipulation of controlled quantum electronic or photonic states is a challenge: electronic quantum optics and quantum electrodynamics of mesoscopic circuits. 

On the one hand, an ideal test-bed for the former is offered by quantum Hall states. There, Coulomb interactions are fundamental to understand the FQHE and the emergence of fractional charges, \cite{fraction_exp} and they couple edge states in the IQHE. Electronic quantum optics is associated with the injection of on-demand electronic excitations and their time evolution through an interacting region. A first theoretical step to address this problematic was initiated by the author \cite{ines_schulz_group,ines_epj} by implementing a scattering approach for plasmon modes with time dependent boundary conditions. This showed charge fractionalisation, \cite{ines_schulz_group,ines_epj,fractionnalisation_eugene_mach_zhender_prb_2008,plasmon_ines_IQHE_HOM_feve_Nature_2015_cite,plasmon_artificial_TLL_experiment_fujisawa_nature_2014_cite} which plays an important role in decoherence \cite{pierre_equilibration_IQHE_Nature_2010,tomography_Grenier_2011} and laid the foundation of NE bosonisation. \cite{out_of_equilibrium_bosonisation_eugene_PRL_2009}
Electronic quantum optics has become an independent field owing to pioneering experimental and theoretical achievements. \cite{quantum_optics_grenier_2011} We mention for instance the analog for electrons of a single photon gun based on a mesoscopic capacitor, \cite{feve_07_on_demand} and implementation of minimal excitations generated by lorentzian pulses. \cite{glattli_levitons_nature_13,klich_levitov,keeling_06_ivanov} In interferometers \cite{sukho_mac_zhender_PRB_2009} such as Hanbury-Brown and Twiss or Hong-Ou-Mandel (HOM) type setups, PASN has offered a tool to explore the charge fractionalisation, \cite{plasmon_ines_IQHE_experiment_feve_2013_cite,plasmon_ines_IQHE_HOM_feve_Nature_2015_cite} to characterize minimal excitations and their statistics \cite{glattli_levitons_nature_13,Sassetti_HOM,Bocquillon_13_splitter_electrons_demand,glattli_levitons_physica_2017,glattli_imen_2022} or to perform electronic tomography \cite{tomography_glattli_2014,tomography_degiovanni_feve_2019}.

On the other hand, quantum electrodynamics of mesoscopic circuits, based for instance on macroscopic atoms such as JJs, requires understanding of radiation-matter interactions, where the radiation corresponds to photons in the electromagnetic environment (for a recent review, see Ref.[\onlinecite{girvin_circuit_quantum_electrodynamics_review_2021}]). Such interactions give rise to the dynamical Coulomb blockade phenomena,\cite{ingold_nazarov} which, in the strong back-action regime, offers a quantum simulation of strongly correlated one dimensional conductors. \cite{ines_saleur,ines_pierre,pierre_anne_boulat_universality} Addressing the statistics of quantum states for both photons and electrons and the generation of squeezed photonic states has been based on finite frequency noise in an ac driven circuit;\cite{squeezing_reulet_PRL_2013,squeezing_reulet_PRL_2015,squeezing_DCB_leppakangas_PRB_2018,squeezing_sasseti_NJP_2021} minimal excitations might offer an interesting basis in this framework. \cite{mora_squeezing_bis} 

It is indeed in an ac driven quantum circuit that some of the NE FRs we have obtained at finite frequencies \cite{ines_cond_mat} have been first tested experimentally.\cite{ines_portier_2015}
 They have been also used to achieve a robust determination of the fractional charge,\cite{glattli_photo_2018,ines_gwendal} or for analyzing experimental investigation of two-particle collisions in a HOM type geometry in the IQHE and the FQHE.\cite{Imen_thesis,glattli_imen_2022} 

The present paper is focused on the PASN at zero frequency, while finite frequency noise is reported to a separate one. Here we present some consequences and applications of the NE FRs for the PASN. We express the PASN in terms of current cumulants of a non-gaussian source, such as a quantum conductor in the classical regime we've studied in Ref.\onlinecite{ines_eugene_detection}. We also derive relations for the PASN's differentials with respect to the ac voltage, then apply them to propose novel methods for charge determination and shot noise spectroscopy. We also derive an important universal inequality, showing that the PASN is super-poissonian. This allows us to state that minimal excitations in non-linear conductors ensure poissonian PASN. We therefore provide an alternative characterization to that by L. Levitov {\it et al}, \cite{keeling_06_ivanov} rather restricted to a linear system. This gives a more thorough analysis than the one we presented in Ref.[\onlinecite{ines_cond_mat}], and which was recovered in the specific model of a TLL.\cite{martin_sassetti_prl_2017} \\
This is the plan of the paper. In section \ref{sec_model}, we recall the family of models and the minimal conditions required by the NE perturbative approach, discussing specifically its validity and limitations for quantum Hall edge states. We derive NE FRs for the PASN and its differentials in section \ref{sec_FR}, and specify to random sources or an initial thermal equilibrium. In section \ref{sec_bounds}, we show that the universal lower bound on the PASN is given by the PAC, and not necessarily by the noise in the dc regime, shown to be higher than the PASN in a SIS junction. This leads us to revisit the criteria for minimal excitations in section \ref{sec_levitov}. We finally discuss (in section \ref{sec_applications}) two other applications based on differentials of the PASN with respect to the ac voltage: shot-noise spectroscopy and determination of the fractional charge.

\begin{center}
\begin{tabular}{|c|c|}
\hline
PAC & Photo-assisted current\\
\hline
PASN & Photo-assisted shot-noise\\
\hline
NE & Non-equilibrium\\
\hline
HOM & Hong-Ou-Mandel \\
\hline
FR & Fluctuation Relation (non-equilibrium) \\
\hline
QPC &Quantum Point Contact\\
\hline
JJ & Josephson Junction\\
\hline
IQHE & Integer Quantum Hall Effect \\
\hline
FQHE & Fractional Quantum Hall Effect\\
\hline
TLL & Tomonaga-Luttinger Liquid\\
\hline
SIN & Superconductor-Insulator-Normal \\
\hline
SIS & Superconductor-Insulator-Superconductor\\
\hline
\end{tabular}
\end{center}
\vspace{1cm}

\section{The perturbative approach}\label{sec_model}
\subsection{The model and minimal conditions}
We consider the Hamiltonian underlying the NE perturbative approach: \cite{ines_eugene,ines_cond_mat,ines_PRB_2019}
\begin{subequations}\label{Hamiltonian} 
\begin{align}
	\mathcal{H}(t)&=\mathcal{H}_0 +e^{-i\omega_Jt} {p}(t){A}+e^{i\omega_Jt}{p}^*(t){A}^{\dagger},
\end{align}
\end{subequations}
where the unperturbed and perturbing terms $\mathcal{H}_0$ and ${A}$ are not specified, nor is the complex function ${p}(t)$, which can be non-periodic, and whose phase $\varphi(t)$ as well as its modulus can depend on time:  \begin{equation}\label{{f}(t)}
{p}(t)\!\! = |{p}(t)| e^{-i{\varphi}(t)}.
\end{equation} 
We adopt the convention that any constant part of a global phase derivative is incorporated into $\omega_J $, so that $\int dt \partial_t\varphi(t)= 0$. 

We focus on transport associated with a given charge operator $\hat{Q}$ assumed to commute with ${\mathcal{H}}_0$ and to be translated through ${A}$ by $e^*$: 
\begin{equation}\label{commutation}[{A},\hat{Q}]=e^* {A},\end{equation} where $e^*$ is a model-dependent charge parameter. Thus the associated current operator reads, in view of Eq.\eqref{Hamiltonian}:
\begin{equation}
	\label{eq:current}
	\hat{I}(t)\!= \partial_t\hat{Q}(t)=
		-i\frac{e^*}{\hbar}\left(e^{-i\omega_Jt}{{{p}}}(t)\;{{A}}- e^{i\omega_Jt}{p}^*(t)\; 
{{A}}^{\dagger}\right)\,.
\end{equation}  
 Other charge operators not conserved by $\mathcal{H}_0$ might enter and couple to other independent constant forces, such as those associated with an electromagnetic environment. 
 \begin{figure}[htbp]
\begin{center}
\includegraphics[width=6cm]{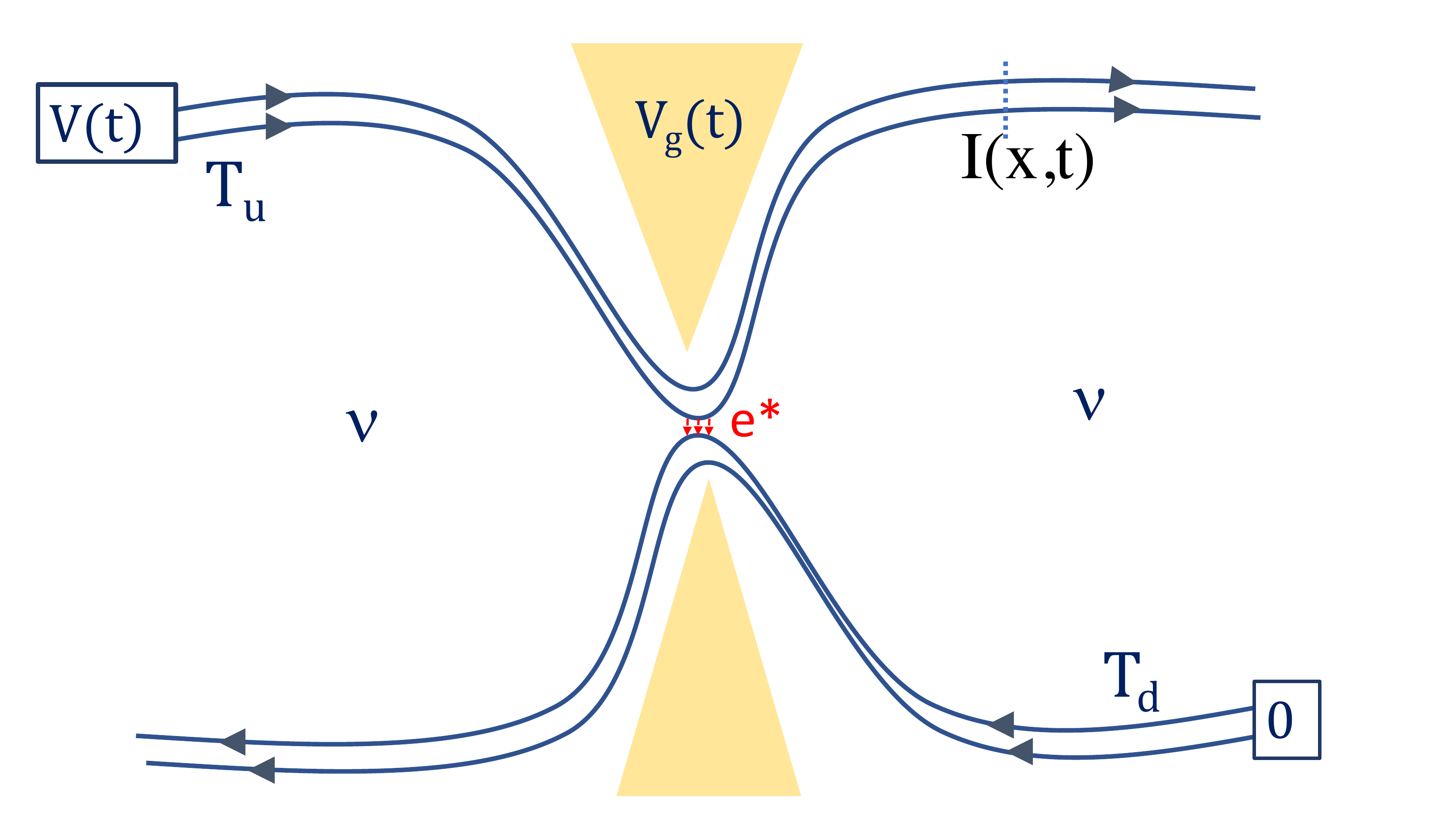}
\caption{\small First example: a QPC in the quantum Hall regime at an integer or fractional filling factor $\nu$. One can include arbitrary profile, range and inhomogeneities of interactions between edge states, as well as abelian or non-abelian statistics. It is possible to have simultaneous time dependence of the voltage reservoirs and the gate, as well as different upper and down temperatures $T_u,T_d$ or imperfect equilibration between edge states. $I(x,t)$ denotes the average chiral current at a position $x$ along the upper edge.}
\label{fig_Hall}
\end{center}
\end{figure}
Indeed the operator $A$ can be a superposition of terms associated with many positions, channels or circuit elements: ${A}= \sum_i\! \!  \;{A}_i,$
or a continuous integral over spatially extended processes. Nonetheless, compared to the dc regime, this generalization is constrained by the fact that all time dependent fields must be incorporated into the single complex function ${p}(t)$.

The main other conditions for the approach are: (i) $A$ is weak, with respect to which second order perturbative theory is valid (ii) only correlators implying ${A}$ and its hermitian conjugate are finite (see Eq.\eqref{Xup_Xdown}). The condition (ii) leads, for a family of distributions $\rho_0$, \cite{ines_PRB_2019} to a vanishing dc current average at $\omega_J=0$; in particular, in superconducting junctions, supercurrent must be negligible due to coupling to a dissipative environment or magnetic fields.

\begin{figure}[htbp]
\begin{center}
\includegraphics[width=6cm]{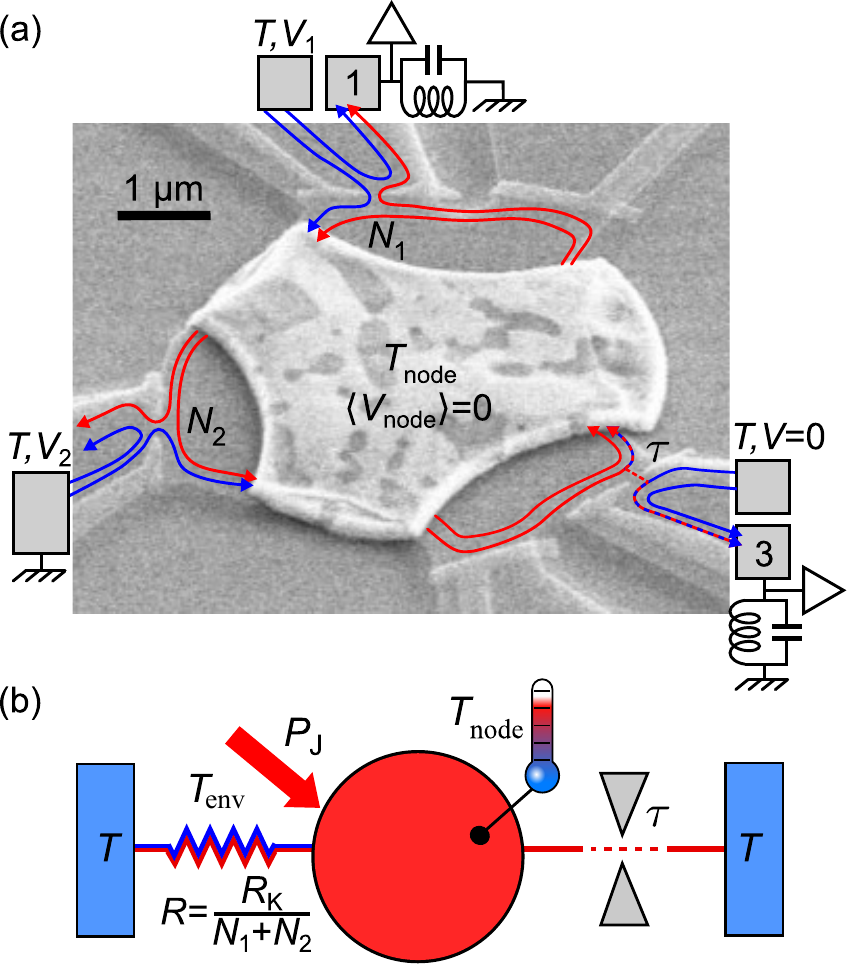}
\caption{\small Second example: a quantum circuit formed by a QPC (on the right side of the lower scheme) coupled to an electromagnetic environment and with a temperature bias, studied in Ref.\cite{ines_pierre_thermal_2021} in the dc regime. This is example of such a quantum circuit studied in Ref.\cite{ines_pierre_thermal_2021} to address dynamical Coulomb blockade. The present NE FR extends to the two opposite conducting and insulating regimes of the quantum phase transition and yields PASN through the QPC in case both the potential drop and gate voltage are time-dependent.}
\label{fig_pierre}
\end{center}
\end{figure}

Interestingly, the approach is not restricted to an initially thermalized system,  \cite{ines_PRB_2019,ines_PRB_R_noise_2020} but extends to an initial stationary NE density matrix ${\rho}_0$ obeying: $[{\rho}_0,\mathcal{H}_0]=0$. Thus $\omega_J$ can be superimposed on other constant independent forces, or one can consider a quantum circuit with a temperature bias \cite{ines_pierre_thermal_2021}(see Fig.\ref{fig_pierre}). 
 
Generically, though not systematically, the coupling to a voltage $V(t)$ can be included into a term $\hat{Q}V(t)$ that can be absorbed  by a unitary transformation \cite{ines_eugene} so that $\omega_J$  (Eq.\eqref{Hamiltonian}) and $\varphi(t)$ (Eq.\eqref{{f}(t)}) obey the following Josephson-type relations, determined by $e^*$ (in view of Eq.\eqref{commutation}):
\begin{subequations}
\begin{align}
\omega_J&=\frac{e^*}{\hbar}{V}_{dc}\label{josephson}\\
\partial_t{\varphi}(t)&=\frac{e^*}{\hbar}{V}_{ac}(t),\label{eq:JR}
\end{align}
\end{subequations}
where $V_{ac}(t),V_{dc}$ are the ac and dc parts of $V(t)$.
But more generally, the common charge $e^*$ could be replaced by two different effective charges, and the above relations can even be broken for NE states, as is the case for the anyon collider. \cite{rosenow,ines_PRB_R_noise_2020,fractionnalisation_anyon_collider_Mora_2021} For generality, we leave $\omega_J$ and ${p}(t)$ (with its amplitude and phase) as unspecified parameters of the model. \cite{note_voltages}

We have previously shown that the average current induced by $p(t)$, $\langle \hat{I}_{\mathcal{H}}(t)\rangle
$, can be, at any time, fully expressed in terms of the dc characteristics only, $I_{dc}(\omega_J)$, whether ${p}(t)$ is periodic \cite{ines_eugene} or not \cite{ines_cond_mat,ines_PRB_2019}. The subscript $\mathcal{H}$ refers to the Heisenberg representation with respect to the Hamiltonian $\mathcal{H}(t)$. In the zero-frequency limit, one gets the PAC: \begin{equation}\label{average_current}I_{ph}(\omega_J)=\int_{-T_0/2}^{T_0/2}\frac{dt}{T_0}\langle \hat{I}_{\mathcal{H}}(t)\rangle,\end{equation}
whose expression will be recalled in Eq.\eqref{Iphoto}. Only dependence on the dc frequency $\omega_J$ is made explicit, while that on $p(t)$ is implicit through the subscript $ph$. Here $T_0$ is the period for periodic $p(t)$, and is a long measurement time for non-periodic $p(t)$ for which it forms the key of a regularization procedure we have proposed. \cite{ines_PRB_2019} We think that this solves the divergency problem obtained in previous works, and compared by H. Lee and L. S. Levitov \cite{lee_levitov} to the orthogonality catastrophe problem, for instance when $V(t)$ is formed by for a single lorentzian pulse. A similar procedure can be carried on for the PASN, defined by:
\begin{equation}\label{Sdefinition_zero}
 {S}_{ph}(\omega_{J})=\int_{-T_0/2}^{T_0/2}\frac{dt}{T_0}\int_{-\infty}^{\infty}  d\tau \left< \delta\hat I_{\mathcal{H}}\left(t-\frac{\tau}2\right)\delta\hat I_{\mathcal{H}}\left(t+\frac{\tau}2\right)\right>,
 \end{equation} 
where $\delta\hat I_{\mathcal{H}}=\hat I_{\mathcal{H}}(t)-\! \langle \hat{I}_{\mathcal{H}}(t)\rangle$.  We will nonetheless simplify it by assuming that the Fourier transform of $p(t)$, $p(\omega)$, is regular at zero frequency, and by referring to Ref.[\onlinecite{ines_PRB_2019}] if not. We will show that $S_{ph}(\omega_J)$ is determined, through a universal FR given by Eq.\eqref{Sphoto}, by $S_{dc}(\omega_J)$, the NE shot noise in the dc regime (which will be coined as the dc noise). It is only when the initial density matrix is thermal that $S_{dc}(\omega_J)$ is determined by $I_{dc}$, and so is the PASN.

\begin{figure}[bp]
\begin{center}
\includegraphics[width=6cm]{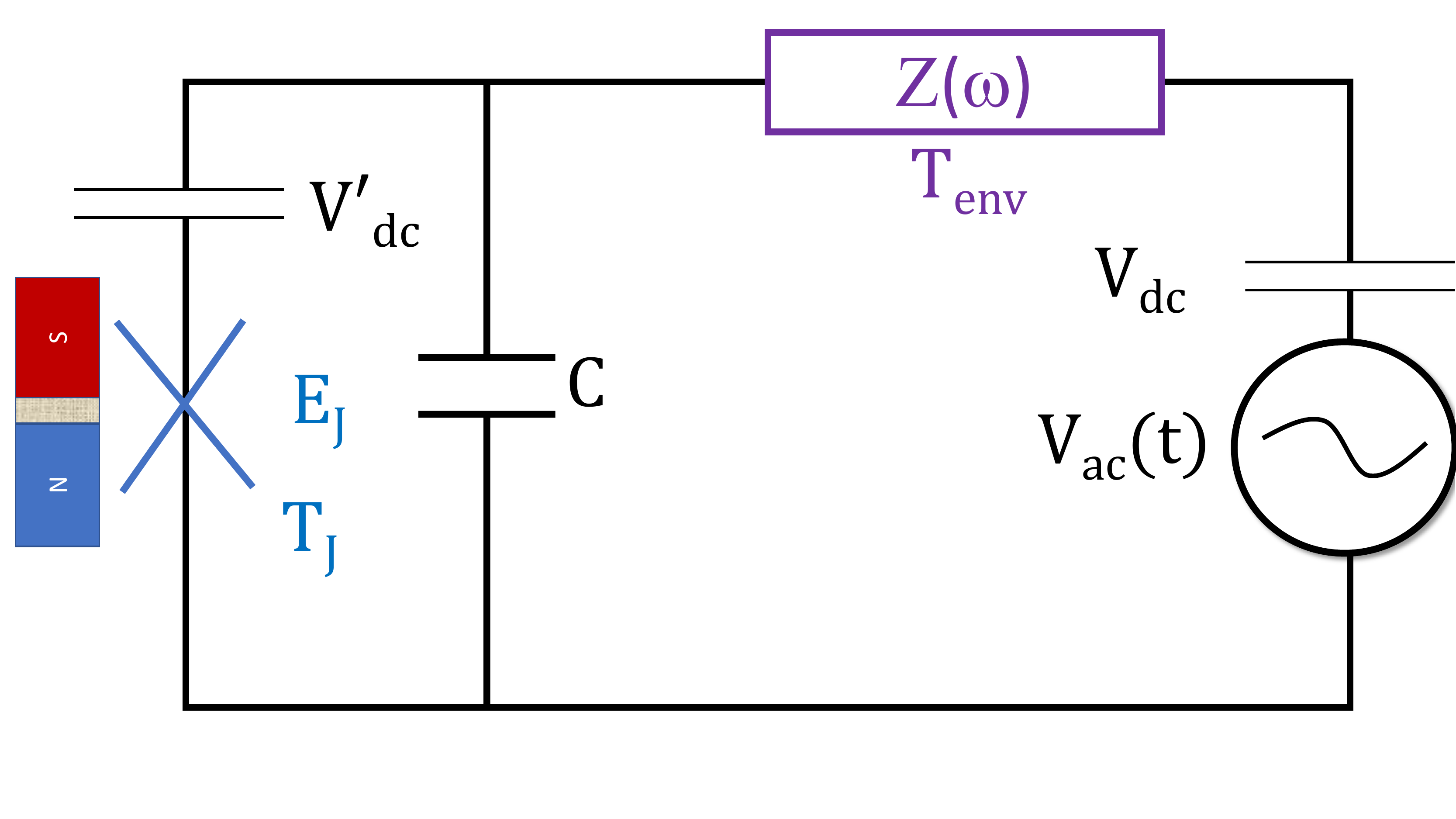}
\caption{\small Third example: a JJ with a small Josephson energy $E_J$ or a NIS junction strongly coupled to an electromagnetic environment. An additional dc voltage $V'_{dc}$ can enter into the Hamiltonian in Eq.\eqref{Hamiltonian} or in the NE stationary density matrix $\rho_0$. }
\label{fig_JJ}
\end{center}
\end{figure}

Some examples of models for which these relations hold are detailed in Ref.[\onlinecite{ines_PRB_2019}] and are illustrated in Figs.(\ref{fig_Hall},\ref{fig_pierre},\ref{fig_JJ}). For instance, $\hat{I}(t)$ is a tunneling current in case $A$ refers to a tunneling term between strongly correlated conductors with mutual Coulomb interactions. It is the Josephson current in a JJ at energies below the superconducting gap $\Delta$ (Fig.\ref{fig_JJ}), for which one has $e^*=2e$. 
Let's discuss in more details the validity and limitations of the approach for a QPC in the quantum Hall regime.

\subsection{Validity of the approach in quantum Hall states}
\label{subsec_Hall}
For a QPC in the FQHE or IQHE at a filling factor $\nu$, the perturbative approach applies to two opposite regimes: the weak backscattering one (when the QPC is almost open, see Fig.(\ref{fig_Hall})), where $\hat{I}(t)$ in Eq.\eqref{eq:current} is a backscattering current, and the strong backscattering regime (when the QPC is pinched off), where $\hat{I}(t)$ is an electron tunneling current. While one has $e^*=e$ in the latter regime, one expects $e^*/e$ to be a fraction in the former when one deals with the FQHE. Many theoretical approaches are based on effective bosonized theories, such as the chiral TLL description for interacting edges in the IQHE or Laughlin series in the FQHE given by $\nu=1/(2n+1)$ with integer $n$, for which $e^*=\nu e$. For other $\nu$ belonging to hierarchical series, there might be many possible models\cite{cheianov_tunnel_FQHE_2015} leading to different values of the dominant charge $e^*$ (that for which the quasiparticle field has the smallest scaling dimension $\delta$. \cite{note_scaling}) It is also frequent that two or more different quasi-particle fields with the same charge and dimension enter into $A$, a situation to which the approach can still be adapted.\\ Let us focus on the case where the QPC is almost open. Such effective theories predict a power law behavior and a crossover energy scale $k_BT_B$ below which the strong backscattering regime is reached, leading to a vanishing dc conductance when both voltages and temperatures vanish. Thus when one adopts effective theories, this delimits the validity of the perturbative approach in both regimes with respect to $T_B$. 

Nonetheless, in experimental works aiming to determine fractional charge \cite{saminad,glattli_photo_2018,ines_gwendal} and statistics, \cite{fractional_statistics_gwendal_science_2020} the measured dc current is not in accordance with this power law behavior. 
Our approach has the advantage to be valid without a specific Hamiltonian nor voltage dependence of the dc current. This explains why the NE FRs we obtained \cite{ines_cond,ines_degiovanni_2016} provided robust methods to determine $e^*=e/5$ at $\nu=2/5$ in Ref.[\onlinecite{glattli_photo_2018}] and $e^*=e/3$ at $\nu=2/3$ in Ref.[\onlinecite{ines_gwendal}].

Though bosonization is not even necessary for the Hamiltonian in Eq.\eqref{Hamiltonian}, one might require, to end up with this form, additional conditions. For instance, absorption of inhomogeneous couplings to ac sources into the function ${p(t)}$ might require that $\mathcal{H}_0$ is a quadratic functional of bosonic fields (not required in the dc regime). 

In order to implement such couplings, one might exploit a useful framework we have initiated \cite{ines_schulz_group,ines_epj}, and which has been largely adopted in electronic quantum optics.\cite{fractionnalisation_IQHE_degiovanni_13,plasmon_ines_IQHE_HOM_feve_Nature_2015_cite} It describes the electronic charge propagation in terms of plasmon dynamics dictated by Coulomb interactions, inducing charge fractionalisation. By developing the equation of motion method for bosonic fields, dynamics is solved for given time dependent boundary conditions dictated by the sources. On the one hand, a classical ac source injects a classical plasmon wave whose time evolution is determined through a scattering matrix for plasmon modes, providing the ac outgoing electronic currents. On the other hand, for a non-gaussian source, such as another QPC different from the central one (replacing the voltage source in Fig.(\ref{fig_Hall})), the NE bosonisation in Ref.\onlinecite{ines_epj} has been extended to take into account statistical fluctuations of the injected current. \cite{out_of_equilibrium_bosonisation_eugene_PRL_2009} Our present NE approach applies to such non-gaussian sources in the dc regime,  \cite{ines_PRB_R_noise_2020} and it is plausible that one can still end up with Eq.\eqref{Hamiltonian} for ac voltages, as we allow for a NE stationary density matrix and a time dependent modulus of ${p}(t)$ that could incorporate ac boundary conditions. For a more rigorous justification and determination of $p(t)$, one needs to combine our treatment of ac voltages \cite{ines_epj} with that of dc non-gaussian sources, \cite{out_of_equilibrium_bosonisation_eugene_PRL_2009} a step not yet achieved to our knowledge.

 At a point $x$ along the edge (Fig.\ref{fig_Hall}), the backscattering average current $\langle \hat{I}_{\mathcal{H}}(t)\rangle$ (see Eqs.\eqref{eq:current},\eqref{average_current}) reduces the perfect linear chiral current in the upper edge (for a derivation in the dc regime, see Ref.\onlinecite{dolcini_05}): 
 \begin{equation}\label{Itotal}
     I(x,t)=  \nu \frac{e^2}hV(t)-\theta(x)\int dt' \lambda(x,t-t')\langle \hat{I}_{\mathcal{H}}(t')\rangle,
 \end{equation}
where $\theta(x)$ is the Heaviside function if the QPC is located at $x=0$. The function $\lambda(x,t)$ is determined by $\mathcal{H}_0$, and describes chiral plasmonic propagation between the QPC and $x$. Denoting its zero-frequency limit by $\lambda$, one gets $I(x,\omega=0)=  \nu {e^2}/h V_{dc}- \theta(x)\lambda I_{ph}(\omega_J).$ One expects $\lambda=\nu$ for simple fractions, but it could be renormalized by non-universal features such as edge reconstruction \cite{halperin_HBT_FQHE_2016}.

It is frequent that one measures rather correlations or cross-correlations between chiral currents, which contain supplementary terms, similarly to Refs.[\onlinecite{trauz_2004},\onlinecite{ines_bena},\onlinecite{ines_bena_crepieux}] in the dc regime. This is also the case when sources are formed by additional QPCs, such as the anyon collider studied in the dc regime; \cite{rosenow,fractional_statistics_gwendal_science_2020,ines_PRB_R_noise_2020,fractionnalisation_anyon_collider_Mora_2021}
 application of ac voltages with a time delay would form a HOM interferometer, as suggested in Ref.[\onlinecite{glattli_levitons_physica_2017}]. It turns out that the perturbative approach is still useful for the supplementary terms, as will be addressed in future works.\cite{Imen_ines}

 \section{Universal fluctuation relations} 
 \label{sec_FR}
Here we first derive the central NE FR for the PASN in Eq.\eqref{Sdefinition_zero}, then apply it to non-gaussian random sources, and finally deduce FRs for the differentials of the PASN with respect to the ac phase, which we  will exploit for the other applications in section \ref{sec_applications}. 
 \subsection{Fluctuation relations between the ac and dc driven regimes}
  
The derivation of the NE FR follows two steps, detailed in Appendix B. The first one yields a second order perturbative expression in terms of two correlators (see Eq.\eqref{Xup_Xdown}) which are evaluated with the Hamiltonian $\mathcal{H}_0$ and the initial NE density matrix ${\rho}_0$, so that they depend only on the time difference $\tau$. Their Fourier transforms at $\omega_J$, denoted by $I_{\rightarrow}(\omega_J),I_{\leftarrow}(\omega_J)$, correspond to dc average currents in two opposite directions induced by $\omega_J$, and determine average current and noise in the dc regime: \begin{subequations}
\begin{align}
I_{dc}(\omega_{\mathrm{J}})&	=
I_{\rightarrow}(\omega_{\mathrm{J}})-I_{\leftarrow}(\omega_{\mathrm{J}})\,
	\label{formal_average_I}\\
S_{dc}(\omega_J)/e^* &=
I_{\rightarrow}(\omega_J)+I_{\leftarrow}(\omega_J)\,.\label{noise_DC_initial_I}
\end{align}
\end{subequations}
Notice that the NE noise $S_{dc}(\omega_J)$ is given by ${S}_{ph}(\omega_{J})$ in Eq.\eqref{Sdefinition_zero} whenever ${p}(t)=1$ in Eq.\eqref{Hamiltonian}. In general, $I_{\rightarrow}(\omega_{\mathrm{J}})\neq I_{\leftarrow}(-\omega_J)$, thus one hasn't necessarily an odd dc current nor an even dc noise.

The second step consists into reversing the two above expressions, so that, alternatively, only the two functions $I_{dc}(\omega_{\mathrm{J}}),S_{dc}(\omega_J)$ determine completely time dependent transport. 
In particular, we can show that the PASN in Eq.\eqref{Sdefinition_zero} is fully determined by $S_{dc}(\omega_J)$ in Eq.\eqref{noise_DC_initial_I}, 
\begin{equation}
S_{ph}(\omega_J)=\int_{-\infty}^{\infty}
\frac{d\omega'}{\Omega_0}\bar{P}(\omega') S_{dc}(\omega'+\omega_J),\label{Sphoto}
\end{equation}
where $\bar{P}(\omega)=|{p}(\omega)|^2$ and $\Omega_0=2\pi/T_0$. 
One recovers the dc regime when ${p}(\omega)=\delta(\omega)$.

  Thus we obtain a universal FR between the ac and dc regimes, which, to our knowledge, has not been derived so far within the present large context of strongly correlated circuits and NE initial states. 
The PASN is a superposition of the noise evaluated at effective dc voltages $\omega_J+\omega'$ for all finite frequencies $\omega'$ of the driving photons, modulated by $\bar{P}(\omega')$. Even at $\omega_{J}=0$, the PASN is determined by the NE dc noise $S_{dc}(\omega')$ (indeed even $S_{dc}(\omega'=0)$ is a NE noise for initial NE states). The above NE FR is independent on the form, range and force of Coulomb interactions or strong coupling to an electromagnetic environment which enters only through the NE dc noise. The external ac or classical noise sources enter through $\bar{P}(\omega')$, which can be viewed as the transfer rate for the many-body eigenstates of $\mathcal{H}_0$ to exchange an energy $\hbar \omega'$ with the ac sources, as can be checked through a spectral decomposition. \cite{ines_philippe_group} 
 
 Experimentally, one gets rid of undesirable contributions by considering the excess PASN. Here we define it by substracting the dc noise in presence of the same dc voltage, $S_{dc}(\omega_J)$, obtained when one switches off the ac source:
\begin{equation}\label{second_choice}
\underline{\Delta} S_{ph}(\omega_J)=S_{ph}(\omega_J)-S_{dc}(\omega_J).
\end{equation}
Let's notice already that $\underline{\Delta} S_{ph}(\omega_J)$ was shown to be always positive by L. Levitov {\it et al} \cite{keeling_06_ivanov} (see Eq.\eqref{levitov_inequality}), but this is not the case in a non-linear SIS junction (as shown in subsection \ref{sec_bounds}), leading us to revisit minimal excitations in section \ref{sec_levitov}.

 Let us now specify to a periodic $p(t)$ with a frequency $\Omega_0$ (see appendix \ref{app_periodic} for more details): \begin{equation}\label{FDT2_periodic_zero}
{S}_{ph}(\omega_{J})= \sum _{l=-\infty}^{+\infty} P_l   S_{dc}(\omega_{J}+l\Omega_0).\end{equation} 
Here $P_l=\bar{P}(l\Omega_0)$ are the transfer rates for many-body states to exchange $l$ photons with the source. It is only when $|p(t)|=1$ that $P_l$ are probabilities, as $\sum_l P_l =1$ (see Eq.\eqref{orthogonality}). 

In case of an initial thermal density matrix $\rho_0  \propto e^{-\beta H_0}$ at a temperature $T=1/\beta$ (see Eq.\eqref{Hamiltonian}), the dc noise obeys the general relation, valid even when $I_{dc}(\omega_J)\neq -I_{dc}(-\omega_J)$: \cite{ines_cond_mat,ines_degiovanni_2016,ines_PRB_R_noise_2020,levitov_reznikov}
\begin{equation}\label{poisson_T}
S_{dc}(\omega_J)=e^*\coth\left(\frac {\hbar\omega_J} {2k_BT}\right) I_{dc}(\omega_J),
   \end{equation}
The PASN is than detailed in appendix \ref{sec_thermal}. Here, focussing on a periodic $p(t)$, on locking values $\omega_J=N\Omega_0$ with an integer $N$ and for $\Omega_0\gg k_B T/\hbar$, we get, from Eq.\eqref{FDT2_periodic_zero}:
\begin{equation}\label{FDT2_periodic_zero_thermal_n}
{S}_{ph}(N\Omega_0)=\sum_{l\neq -N} P_l \;|I_{dc}\left[(N+l)\Omega_0\right]|+2 P_{-N} k_B T G_{dc}(T).\end{equation}
Here $G_{dc}(T)=dI_{dc}(\omega_J)/dV_{dc}$ at $e^*V_{dc}\ll k_BT $; we make explicit its temperature dependence, generic in non-linear systems, while it is implicit in the NE current average and the PASN. Thus we get a mixture between NE and thermal contributions (see appendix \ref{sec_thermal}), similarly to NE finite-frequency noise in the dc regime. \cite{ines_degiovanni_2016} Taking the excess noise in Eq.\eqref{second_choice} does not cancel the thermal one, even though we are in the NE quantum regime. 

This relation unifies and goes beyond previous works restricted to $|p(t)|=1$ and to independent electrons scattered by a linear QPC \cite{dubois_minimization_integer,glattli_levitons_nature_13} or to a TLL in Ref.\onlinecite{crepieux_photo}. 
In the TLL or more general effective theories for the FQHE, it allows us to localize and regularize a divergency noted in the latter work, as will be explained elsewhere. \cite{Imen_ines}

Thus the universal relations in Eqs.\eqref{Sphoto},\eqref{FDT2_periodic_zero} extend fully the lateral-band transmission for the PASN to time dependent tunneling amplitudes and periodic, non-periodic or fluctuating sources. In the large family of strongly correlated circuits these relations cover, NE many-body states replace thermal one-electron states. They are also suited to address two-particle collisions in a symmetric or asymmetric HOM type geometry where two ac sources, periodic or not, operate with a time delay (as noticed briefly in Ref.\onlinecite{ines_PRB_2019}). It has been used in a recent experimental analysis of two-electron collisions \cite{Imen_thesis,glattli_imen_2022} in chiral quantum Hall edges.
\subsection{Fluctuating sources}
One advantage of considering non-periodic $p(t)$ is that one can deal with classical states of radiations. Indeed, if we assume that $|{p}(t)|=1$, one has $\int d\omega' \bar{P}(\omega')/\Omega_0=1$, so that $\bar{P}(\omega)$ becomes a probability.\cite{ines_PRB_2019} It plays a similar role to the $P(E)$ function which yields the probability for a tunneling electron to exchange photons at a frequency $\omega=E/\hbar$ with an electromagnetic environment \cite{ingold_nazarov}. Indeed, this is precisely the meaning of $\bar{P}(\omega)$ if $\varphi(t)$ is associated with a gaussian or non-gaussian electromagnetic environment in the classical limit, formed for instance by a quantum conductor we've studied in Ref.[\onlinecite{ines_eugene_detection}].

More generally, if classical fluctuations of $\varphi(t)$ have a distribution $\mathcal{D}(\varphi)$, one has to take into account averages over $\mathcal{D}(\varphi)$, denoted by $<...>_{\mathcal{D}}$:
\begin{equation}\label{p_average}
\bar{P}(\omega)=\int_{0}^{T_0}\frac{dt}{T_0} e^{i\omega t} <e^{i(\varphi(t)-\varphi(0))}>_{\mathcal{D}}.
\end{equation} 
Notice that we assume here the stationarity of the distribution for $\varphi$ so that $<e^{i(\varphi(t)-\varphi(t'))}>_{\mathcal{D}}$ depends only on $t-t'$, thus dropping the integral over $t+t'$.  
One can further write $<e^{i(\varphi(t)-\varphi(0))}>_{\mathcal{D}}$ as an exponential of cumulants of $\varphi(t)$ at order $m$ ($m$ is an integer; see Ref.\onlinecite{sukho_detection} for the full expression): \begin{eqnarray}
J_m(t)&=&\frac {1} {{m}\mathpunct{!}} <(\varphi(t)-\varphi(0))^m>_{\mathcal{D}}.\label{Jm}
\end{eqnarray}
If we expand it up to $m=3$, justified in the limit of weak coupling, we obtain:
\begin{equation}
S_{ph}(\omega_J)=\iint \frac{d\omega dt}{\Omega_0} S_{dc}(\omega+\omega_J)e^{i\omega t-J_2(t)+iJ_3(t)},\label{Sphoto_cumulant}
\end{equation}
 There might be various ways, depending on regimes and setups, to exploit this link, in particular to use the PASN as a way of detection of cumulants of the quantum conductor, as done with the PAC.\cite{ines_PRB_2019,ines_eugene} Compared to previous works proposing Tunnel junctions or JJs as cumulant detectors,\cite{sukho_detection_JJ_PRL2007,sukho_detection} the present model opens the path to exploit a larger family of strongly correlated detectors which are not necessarily disconnected, and to drive both the detector and the non-gaussian source in stationary NE states. 
 
We insist nonetheless that a quantum environmental phase operator $\hat{\varphi}(t)$ whose dynamics is dictated by the Hamiltonian $\mathcal{H}_0$ can be also encoded into ${A}$ through $e^{i\hat{\varphi}(t)}$, whose correlations affect the PASN through the dc noise $S_{dc}$ according to Eq.\eqref{Sphoto}.

\subsection{Fluctuation relations for differentials of the PASN}
An alternative to excess noise, in order to get rid of undesirable noisy sources, is to consider the derivative of the PASN with respect to the dc voltage, which, in view of the FRs in Eqs.\eqref{Sphoto},\eqref{FDT2_periodic_zero}, is determined through the differential dc noise.

It is also interesting, for some potential applications, to differentiate the PASN with respect to the ac components of the voltage, $V_{ac}(\omega)$, or for more generality, $\varphi(\omega)$ (as Eq.\eqref{eq:JR} is not systematic). Given a non-periodic or random $p(t)$, one can show that $\delta p(\omega')/\delta \varphi(\omega)=-ip(\omega'-\omega)$, so that :
\begin{equation}\label{FDT2_periodic_zero_derivative}
\frac{\delta S_{ph}(\omega_{J})}{\delta \varphi(\omega)}=-i\int \frac{d\omega'}{\Omega_0}  p(\omega')p^*(\omega'+\omega)  \left[S_{dc}(\omega_{J}+\omega'+\omega)-S_{dc}(\omega_{J}+\omega')\right].
\end{equation} 

Let us now take a second differential with respect to $\varphi(-\omega)$. Then we are back to the PASN through an interesting closed relation:
\begin{equation}\label{FDT2_periodic_zero_second_derivative}
\frac{\delta ^2 S_{ph}(\omega_{J})}{\delta\varphi(\omega)\delta\varphi(-\omega)}= S_{ph}(\omega_{J}+\omega)+S_{ph}(\omega_{J}-\omega)-2S_{ph}(\omega_{J})
\end{equation} 
A similar relation holds for the PAC in Eq.\eqref{Iphoto},\cite{ines_eugene} as well as for periodic drives, taking ${\delta ^2 S_{ph}(\omega_{J})}/{\delta \varphi_k\varphi_{-k}}$ and $\varphi_k=\varphi(k\Omega_0)$ for any integer $k$. 
We notice however a possible experimental difficulty in case one lacks a precise knowledge of the effective phase (for instance at the level of the QPC in the Hall regime), which makes noise spectroscopy discussed in \ref{subsec_spectroscopy} useful.

It is indeed easier to consider the limit of the stationary regime, defined by $p(t)=1$ thus $p(\omega')=\delta(\omega')$ (thus the limit of a vanishing phase, up to multiple of $2\pi$, and of a unit modulus). Then the first derivative in Eq.\eqref{FDT2_periodic_zero_derivative} vanishes, and one can replace, on the r.h.s. of Eq.\eqref{FDT2_periodic_zero_second_derivative},  $S_{ph}\rightarrow S_{dc}$. In this limit, we can also show that $\delta^2 S_{ph}(\omega_{J})/{\delta\varphi(\omega)\delta\varphi(-\omega')}\rightarrow 0$ for all $\omega'\neq \omega$. 
Thus, assuming furthermore that $|p(t)|=1$ so that only functional dependence with respect to $\varphi(t)$ enters, the excess PASN defined through Eq.\eqref{second_choice} can be expanded through the main quadratic correction due a small ac phase:
\begin{equation}\label{FDT2_periodic_zero_second_derivative_zero_ac}
 \underline{\Delta} S_{ph}(\omega_J)= \int_{-\infty}^{+\infty} d\omega |\varphi(\omega)|^2\left[ S_{dc}(\omega_{J}+\omega)+S_{dc}(\omega_{J}-\omega)-2S_{dc}(\omega_{J})\right]+o(\varphi^4).\end{equation} This universal relation is coherent with the fact that only one or zero photon processes enter at a low ac modulation.
  For periodic drives the integral is replaced by a discrete sum.
  In particular, in case $\varphi(t)=\varphi_{ac}\cos \Omega_0t$ we get:
  \begin{equation}\label{expansion_sine}\underline{\Delta} S_{ph}(\omega_J)/\left[S_{dc}(\omega_{J}+\Omega_0)+S_{dc}(\omega_{J}-\Omega_0)-2S_{dc}(\omega_{J})\right]=\varphi_{ac}^2.\end{equation} 
  
As will be discussed in \ref{subsec_spectroscopy}, \ref{subsec_fractional_charge}, the above relations offer methods for shot noise spectroscopy and for a robust determination of the fractional charge, based on determining $\omega_J$ and the Josephson-type relation in Eq.\eqref{josephson} whenever it holds. This is in some sense similar to a spectroscopy as $\omega_J$ is determined by external constant forces or voltages. Indeed there are also situations where $\omega_J$ implies other unknown parameters which can then be determined consequently. Two have been addressed in Ref.\onlinecite{ines_PRB_R_noise_2020}: either $\omega_J$ is linked to a non-universal parameter of fractional statistics that enters in analyzing the anyon collider, or to the voltage drop generated by a temperature bias, which yields the Seebeck coefficient. 

\section{Universal lower bounds on the PASN}
\label{sec_bounds}
This section gives crucial features that will allow us to revisit minimal excitations in section \ref{sec_levitov}. 
We will first show that that the universal lower bound provided by L. Levitov {\it et al}\cite{keeling_06_ivanov} is restricted to linear conductors, by giving the counterexample of a non-linear SIS junction with an initial thermal distribution. Than, considering again a NE initial  distribution, we show that it is rather the PAC which provides a universal lower bound for the PASN,  therefore super-poissonian. 

\subsection{Breakdown of the dc noise bound in a non-linear SIS junction}

\label{subsec_SIS}
In an independent-electron picture, the choice for the excess noise, $\underline{\Delta} S_{ph}(\omega_J)$ in Eq.\eqref{second_choice} is motivated by the fact that it arises from the cloud of electron-hole excitations generated by the ac voltage,\cite{FCS_TD_tunnel_09,dubois_minimization_integer}thus inducing a positive excess noise, $\underline{\Delta} S_{ph}(\omega_J)>0$. Indeed, in a more general framework of strongly correlated systems, the ac voltage was shown to increase the noise through a theorem by L. Levitov {\it et al} \cite{klich_levitov,keeling_06_ivanov}:
\begin{equation}\label{levitov_inequality}
S_{ph}(\omega_J)\geq S_{dc}(\omega_J).\end{equation}
Nonetheless, we show now that adding an ac voltage to a dc one could decrease the PASN in a non-linear SIS junction, so that these inequalities are reversed.

\begin{figure}[htbp]
\begin{center}
\includegraphics[width=8cm]{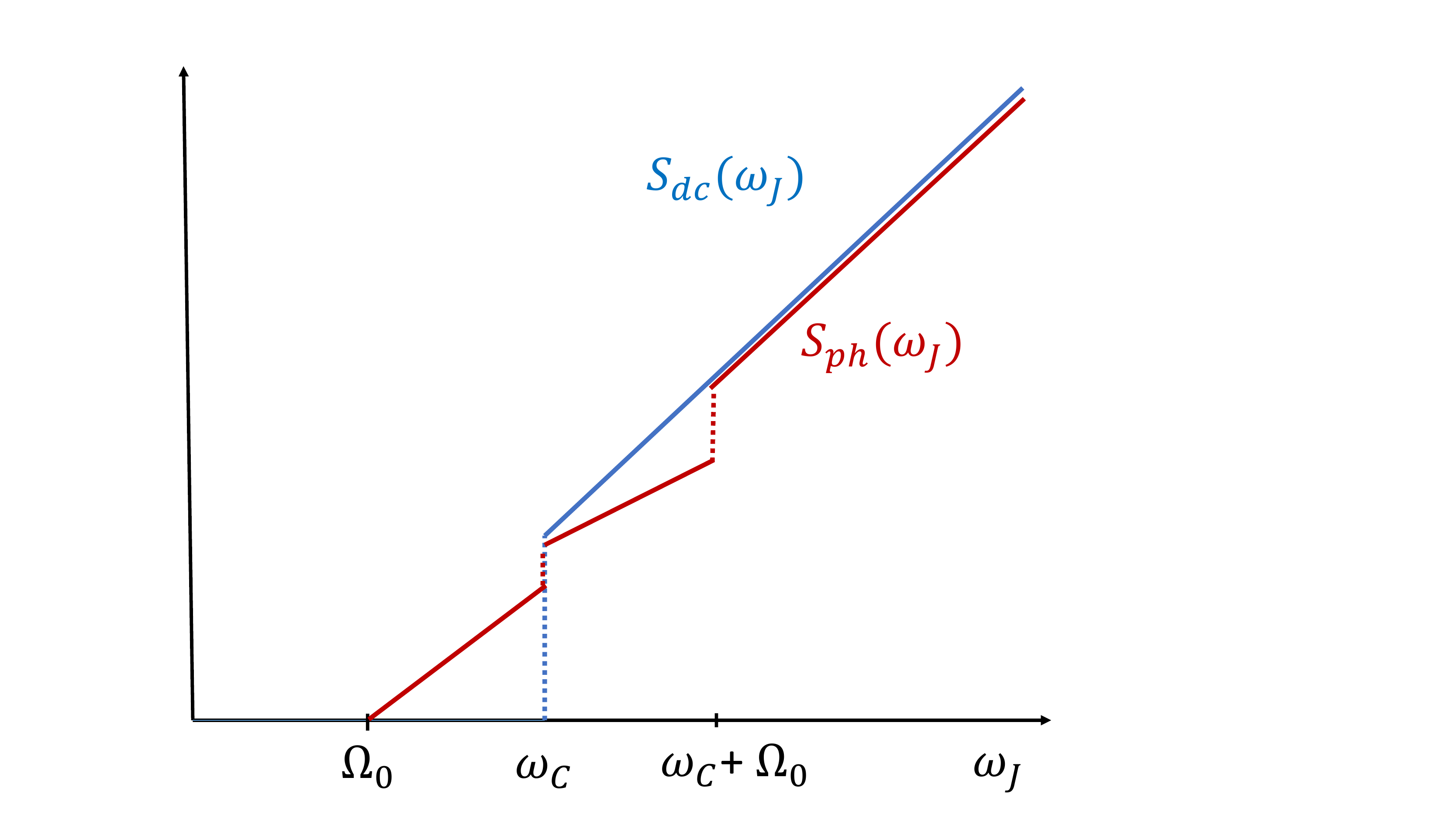}
\caption{\small. The dc noise associated with the quasiparticle current in a SIS junction, as a function of the dc frequency $\omega_J$ (in blue) and the PASN under a small sine voltage (in red). $S_{dc}$ vanishes below a threshold $\omega_c=2\Delta\hbar$ where $\Delta$ is the superconducting gap, and has a linear behavior above, $a\omega_J-b$. The PASN behavior is sketched (thus units are arbitrary) by choosing  $b/a=\Omega_0=\omega_C/2$. It is below $S_{dc}$ at dc voltages above $\omega_C$. }
\label{fig:SIS_I}
\end{center}
\end{figure}

We adopt, in a similar context as these works, an initial thermalized distribution in the zero temperature limit, so that the dc noise is poissonian (see Eq.\eqref{poisson_T}). We also consider a quasi-particle current $I_{dc}$ \cite{tien_gordon,tucker_rev} with a voltage gap $2\Delta/e$, thus a dc frequency gap $\omega_C=2\Delta/\hbar$ (here $e^*=e$), and a linear behavior above: $I_{dc}(\omega_J>0)=\theta(\omega_J-\omega_c)(a\omega_J-b)$ (see Fig.\ref{fig:SIS_I}). 
where $a,b$ are positive coefficients. This gives in particular $G_{dc}(T)=0$. Now we choose the dc and ac frequencies such that $\Omega_0<\omega_C<\omega_J$ and $ b/a<\omega_J-\Omega_0<\omega_C $. We consider a weak enough sine voltage so that we can use the second-order expansion in Eq.\eqref{expansion_sine}. As the dc noise is poissonian, the sign of $\underline{\Delta}S_{ph}$ is that of $I_{dc}(\omega_J+\Omega_0)+I_{dc}(\omega_J-\Omega_0)-2I_{dc}(\omega_J)=a(\Omega_0-\omega_J)+b<0$ (one has $I_{dc}(\omega_J-\Omega_0)=0$ as $\omega_J-\Omega_0<\omega_C$). Therefore the PASN is decreased in this dc voltage range, \begin{equation}
   S_{ph}(\omega_J)< S_{dc}(\omega_J),
\end{equation}
which is at odd with the inequality in Eq.\eqref{levitov_inequality}. We can indeed plot the PASN for all dc voltages (see Fig.\ref{fig:SIS_I}), which is also slightly below the dc noise at $\omega_J>\omega_C+\Omega_0$ for which  $S_{ph}(\omega_J)=(P_0+2P_1)S_{dc}(\omega_J)$. 

Indeed, due to our hypothesis of a weak sine voltage, one has a poissonian PASN for all $\omega_J>\Omega_0$, $S_{ph}(\omega_J)=eI_{ph}(\omega_J)$ (see Eq.\eqref{Iphoto_periodic}). Then our result is coherent with the known fact that $I_{ph}(\omega_J)<I_{dc}(\omega_J)=S_{dc}(\omega_J)/e$ in the range $\omega_C<\omega_J<\omega_C+\Omega_0$. \cite{tucker}
\subsection{Super-poissonian PASN}
Considering again a NE density matrix $\rho_0$ and non-periodic $p(t)$, let us first recall the relation obtained for the PAC in Eq.\eqref{average_current}:
\begin{equation}
I_{ph}(\omega_J)=\int_{-\infty}^{\infty} \frac{d\omega'}{\Omega_0} \bar{P}(\omega') I_{dc}(\omega'+\omega_J).\label{Iphoto}
\end{equation}
Similarly to Eq.\eqref{Sphoto}, it is also interpreted within a lateral-band transmission picture for correlated many-body states. \cite{ines_eugene,ines_PRB_2019}
Now we have shown that the dc noise is super-poissonian\cite{ines_PRB_R_noise_2020} \begin{equation}S_{dc}(\omega_J)\geq e^*|I_{dc}(\omega_J)|\label{S_dc_super}\end{equation} due to the fact that $I_{\rightarrow}(\omega_J)$ and $I_{\leftarrow}(\omega_J)$ are positive (see Eqs.\eqref{formal_average_I},\eqref{noise_DC_initial_I}). This is obviously verified by Eq.\eqref{poisson_T} for an initial thermal equilibrium, which yields a poissonian dc noise at low temperatures. Nonetheless, the super-poissonian dc noise does not arise necessarily from thermal effects if the NE initial distribution is, for instance, generated by additional dc voltages (see Fig.\ref{fig_JJ}) rather than by temperature gradients (as in Fig.\ref{fig_pierre}).

Now by comparing Eq.\eqref{Sphoto} to Eq.\eqref{Iphoto}, and using Eq.\eqref{S_dc_super}, we obtain also a super-poissonian PASN \cite{ines_cond_mat} :
\begin{equation}\label{super_poisson}
{S}_{ph}(\omega_{J})\geq e^*|I_{ph}(\omega_J)|.\end{equation} 
 This is an important inequality, also valid when one has periodic drives, and even when the global system is in the ground state. 
Notice that this inequality suggests an alternative for the excess noise, given by $S_{ph}(\omega_J)-e^*|I_{ph}(\omega_J)|$, which yields always a positive sign, though it is not the most relevant experimentally as discussed in appendix \ref{app_excess}.

Let's now comment on the case the dc current is linear with respect to $\omega_J$ and $|{p}(t)|=1$. Then ${I}_{ph}(\omega_{J})=I_{dc}(\omega_J)$, which becomes linear as well. In particular, if Eqs.\eqref{josephson},\eqref{eq:JR} hold, one has simply:
\begin{equation}\label{photocurrent_linear}
I_{ph}(\omega_J)=G_{dc} V_{dc},\end{equation}
 where $G_{dc}=I_{dc}(\omega_J)/V_{dc}$ is the linear conductance.
 Therefore, the lower bound on the PASN becomes given by the dc current
${S}_{ph}(\omega_{J})\geq e^*|I_{dc}(\omega_J)|$, exactly as is the case for the dc noise in Eq.\eqref{S_dc_super}. Nonetheless, it is only when the system is in the ground state, thus at low temperatures for a thermal equilibrium distribution, that the dc current can be replaced by the dc noise (see Eq.\eqref{poisson_T}), so that one recovers Eq.\eqref{levitov_inequality}.  

The inequality in Eq.\eqref{super_poisson} offers an alternative to Eq.\eqref{levitov_inequality}, valid in the SIS junction we addressed above for all dc voltages. Though restricted to a perturbative regime, it covers a much larger family of non-linear systems and quantum circuits. But an important difference from Eq.\eqref{levitov_inequality} is that the PAC, forming the universal lower bound, is also determined by the ac voltage.
 

\section{Revisiting minimal excitations}
\label{sec_levitov}
In view of the above features, we address the issue of characterizing minimal excitations, whose realization requires a ground  many-body state, for instance the low-temperature limit of an initial thermal equilibrium. 
 \subsection{L. Levitov's characterisation: limitation to linear conductors}
Characterization of minimal excitations (we focus here on "electron" type ones) by L. Levitov {\it et al}\cite{keeling_06_ivanov} through the PASN is based on the central inequality in Eq.\eqref{levitov_inequality}. 

First, the authors imposes an injected charge per period $Q_{cycle}= Ne$. As they assume that $I(t)=\partial_tQ(t)=e^2 V(t)/h$ (for a linear ballistic conductance with non-interacting electrons), one has $Q_{cycle}={e^2}\int _0^{T_0}dtV(t)/h= {e^2}T_0V_{dc}/h$, controlled by the dc component of the voltage $V(t)$ only. This leads to the condition $V_{dc}=Nh/(T_0 e)$ or, taking $e^*=e$ in Eq.\eqref{josephson}, to $\omega_J=N\Omega_0$. 

Secondly, according to Eq.\eqref{levitov_inequality}, the voltage which minimizes the PASN by injecting well-defined electronic excitations must ensure the equality $S_{ph}(\omega_J)= S_{dc}(\omega_J)$, the lower bound of the PASN. This requires that the Fourier components $p_l$ of ${p}(t)=e^{-i\varphi(t)}$ obey:
\begin{equation}\label{p_l_lorentz}
    l<-N\implies p_l=0.
\end{equation}For that, the total voltage must be formed by a series of lorentzian pulses centered at $kT_0$ with a width $2W$, so that the phase derivative verifies (see Eq.\eqref{eq:JR}, thus $\int dt  \partial_t\varphi{(t)}=0$):
\begin{equation}\label{V_lorentzian}
\partial_t\varphi(t)=\frac{N\Omega_0}{\pi}\sum_{k=-\infty}^{\infty}\frac{1} {1+(t-kT_0)^2/W^2}-N\Omega_0.
\end{equation} 
Nonetheless, such a characterisation require the current to be linear, thus doesn't apply to a QPC in the FQHE with a non-linear dc current as claimed in Ref.[\onlinecite{keeling_06_ivanov}]. 

Let us give three reasons for that. 
First, the injected charge corresponds to the PAC  in Eq.\eqref{Iphoto}, which, for a non-linear dc current, has a non-trivial functional dependence on the ac voltage \cite{ines_PRB_2019}.

Second, let us adopt the lorentzian pulses, and apply Eq.\eqref{p_l_lorentz} to the FR in Eq.\eqref{FDT2_periodic_zero} :
\begin{subequations}
\begin{align}
\label{S_frequency_S_dc_zero_lorentzian}
{S}_{ph}(N\Omega_0)&=\sum_{l\geq -N }^{\infty} P_l S_{dc}((N+l)\Omega_0),
  \end{align}
   \end{subequations}
The equality ${S}_{ph}(\omega_J)=S_{dc}(\omega_J)$, given a dc poissonian noise, requires in general a linear dc current (notice that one has to add $2k_BG_{dc}(T)TP_{-N}$ on its r.h.s., in view of Eq.\eqref{FDT2_periodic_zero_thermal_n}). 

Third, the authors were not aware of an implicit hypothesis underlying the inequality in Eq.\eqref{levitov_inequality}, that of the dc current must be linear. So it cannot be generalized to  non-linear conductors, such as the SIS junction we considered in subsection \ref{subsec_SIS}.

\subsection{Super-Poissonian to poissonian PASN: minimal excitations}
 We have shown that the PASN is universally super-poissonian, whatever is the initial NE  in Eq.\eqref{super_poisson}. This is a first central ingredient of our alternative path. The second one is to define minimal excitations as those for which the PASN becomes poissonian, thus equality is reached in Eq.\eqref{super_poisson}. For that we need to specify to an initial many body ground state. We focus, for simplicity, on a periodic $p(t)$ with $|p(t)|=1$.
 
 Instead of solving for the voltage, we gain generality by reasoning in terms of $\varphi(t)$ and the dc frequency $\omega_J$ (the relations in Eqs.\eqref{josephson},\eqref{eq:JR} are not systematic). Now $\omega_J$, which doesn't fix the transferred charge, is not fixed but has rather to be determined, on the same level as $\varphi({t})$, by requiring equality in Eq.\eqref{super_poisson}. For that, we write Eqs.\eqref{FDT2_periodic_zero_thermal},\eqref{Iphoto_periodic} in the limit of a strictly zero temperature:
\begin{subequations}
\begin{align}
\label{S_I_I_+_I_-}
{S}_{ph}(\omega_J)&=e^*[I_{ph,+}(\omega_J)-I_{ph,-}(\omega_J)].\\
{I}_{ph}(\omega_J)&=I_{ph,+}(\omega_J)+I_{ph,-}(\omega_J).
  \end{align}
   \end{subequations}
We have separated
$
{I}_{ph,\pm}(\omega_J)=\sum_{\pm(\omega_J+l\Omega_0) \geq 0} {P}_l \;I_{dc}(\omega_J+l\Omega_0)$, the contributions to the PAC generated by either positive or negative effective dc drives. We have used the fact that $I_{dc}(\omega_J=0)=0$ and $\omega_J I_{dc}(\omega_J)\geq 0$  for a thermal distribution, \cite{ines_PRB_2019} so that
 $\pm{I}_{ph,\pm}(\omega_J)\geq 0$. 
Therefore the poissonian limit is reached whenever $I_{ph,+}(\omega_J)=0$ or $I_{ph,-}(\omega_J)=0$. We focus here on the condition $I_{ph,-}(\omega_J)=0$. If it has to be ensured whatever the profile of $I_{dc}$, it requires that ${P}_l=0$ for all $l$ such that $\omega_J+l\Omega_0<0$. Then one can show, using similar arguments to those by L. Levitov {\it et al},\cite{keeling_06_ivanov} that the phase must have the form in Eq.\eqref{V_lorentzian}, and that $\omega_J=N\Omega_0$ due to analytic properties of ${p}(t)$ in the complex plane.

Therefore, we get, from Eq.\eqref{S_frequency_S_dc_zero_lorentzian}:
\begin{equation}{S}_{ph}(N\Omega_0)=e^*|I_{ph}(N\Omega_0)|.\label{poisson_lorentz}\end{equation} 
 This poissonian regime indicates that the PASN reduces to the average charge given by $e^*|I_{ph}(N\Omega_0)|$, now generated only by photon absorption of the many-body ground state. Indeed, since temperatures are always finite, and even for the present NE quantum regime with $T\ll \hbar\Omega_0/k_B$, one has: $S_{ph}(N\Omega_0)=e^*|I_{ph}(N\Omega_0)|+2k_B TP_{-N}G_{dc}(T)$ (see Eq.\eqref{FDT2_periodic_zero_thermal_n}).
 
Similarly, in case one superimposes a finite dc frequency $\omega_{dc}$ on top of $N\Omega_0$, one goes back to a super-poissonnian PASN. Let's give an example for $N=1$ and decrease the dc drive by a frequency $\omega_{dc}$ verifying $k_BT /\hbar \ll \omega_{dc}<\Omega_0$. Then we get: $${S}_{ph}(\Omega_0-\omega_{dc})=e^*|I_{ph}(\Omega_0-\omega_{dc})|+2e^*P_{-1} |I_{dc}(-\omega_{dc})|.$$ 

This analysis provides another example at odd with Eq.\eqref{levitov_inequality}, by considering again a SIS junction in the ground state (see Fig.(\ref{fig:SIS_I})). As mentioned in subsection \ref{subsec_SIS}, a sine voltage reduces the PAC in the range $\omega_C<\omega_J<\omega_C+\Omega_0$ compared to $I_{dc}(\omega_J)$, \cite{tucker} a result one can extend to an arbitrary profile of the voltage. Since we showed that lorentzian pulses generate a poissonian PASN (we don't have any thermal contribution as $G_{dc}(T)=0$), one has ${S}_{ph}(N\Omega_0)<S_{dc}(N\Omega_0)$ if $N$ verifies $0<N\Omega_0-\omega_C<\Omega_0$. 

Notice also that the poissonian limit can be reached by a weak sine voltage applied to the SIS junction, thus is not exclusive to lorentzian pulses.

Our analysis can be extended to a non-periodic $p(t)$ with a possible time-dependent modulus $|{p}(t)|$, where similar analytic properties of ${p}(\omega)$ lead to a poissonian PASN. 

We finally insist that for a NE initial distribution, the inequality in Eq.\eqref{super_poisson} remains strict even for Lorentzian pulses.

\subsection{FQHE: non-trivial charge of minimal excitations and superpoissonian PASN}
 In the FQHE, the renormalization by a fractional charge $e^*$ in front of the current arises from the fact that $A$ translates the charge by $e^*$, which is chosen as the dominant process with the lower dimension $\delta$. But contrary to the initial claim of L. Levitov {\it et al}, the lorentzian pulses cannot carry $Ne^*$ per cycle. It was shown, in Refs.\onlinecite{glattli_levitons_physica_2017,martin_sassetti_prl_2017}, that one has still $Q_{cycle}=Ne$. Nonetheless, the given proof was restricted to Laughlin states, $\nu=1/(2n+1)$, for which $e^*=\nu e$.\\
Let's consider hierarchical states, such as the Jain series in the recent experimental works (probing $e^*=e/5$ at $\nu=2/5$ in Ref.\onlinecite{glattli_photo_2018} or $e^*=e/3$ at $\nu=2/3$ in Ref.\onlinecite{ines_gwendal}). One needs to assume in order to reach almost poissonian PASN, that the lorentzian pulses are not deformed at the level of the QPC. We also assume that Eqs.\eqref{josephson},\eqref{eq:JR} hold, where $e^*$ enters, so that the condition $\omega_J=N\Omega_0$ means that the value of $V_{dc}$ for a given frequency $\Omega_0$ depends on $e^*$.

Given this condition, we would like to provide the charge carried by a minimal injected excitation in the region before the QPC, where the chiral current reduces to the first term on the r.h.s. of Eq.\eqref{Itotal}, thus for $x<0$. The charge per cycle is given by (as $\omega_J=N\Omega_0=2N\pi /T_0$):
 \begin{equation}
 Q_{cycle}=\nu \frac{e^2}h\int _0^{T_0}dtV(t) =Ne\frac{\nu e }{e^*}.
 \end{equation}
This suggests that $Q_{cycle}$ gives a possible access to $e^*$, as one generally determines $\nu$ from conductance plateaus. Since $\nu$ is not a simple fraction, the bosonisation allows for many models whose dominant backscattering process (with the smallest scaling dimension \cite{note_scaling}) can carry different charges $e^*$. For instance, for $\nu=2/(2n+1)$ with integer $n$, some models lead to $e^*=e/(2n+1)$,\cite{cheianov_tunnel_FQHE_2015} so that $Q_{cycle}=2Ne$ is integer, which is the same charge as that in the IQHE at $\nu=2$. 
  
Now consider the weak backscattering regime, at energies above $k_BT_B$. In case one uses the effective theories leading to power law behavior, the thermal contribution to the PASN in Eq.\eqref{FDT2_periodic_zero_thermal_n} cannot be ignored. Therefore we get a strictly super-poissonian PASN, contrary to the claim in Ref.\onlinecite{martin_sassetti_prl_2017}: $S_{ph}(N\Omega_0)=e^*|I_{ph}(N\Omega_0)|+2k_B TP_{-N}G_{dc}(T)$, where $G_{dc}(T)$ is a power of $T$. A more detailed comparison between the two terms will be treated separately.\cite{Imen_ines} 
\section{Other applications}
\label{sec_applications}

\subsection{Shot-noise spectroscopy}
\label{subsec_spectroscopy} 
In general, the transfer rates $\bar{P}(\omega)$ in Eq.\eqref{Sphoto}  might be unknown as they can be affected, for instance, by interactions or by NE or fluctuating sources. Thus one possible advantage of the FR in Eq.\eqref{Sphoto} would reside in shot-noise spectroscopy. A protocol discussed in Ref.\onlinecite{glattli_degiovanni_PRB_2013} is nonetheless restricted to non-interacting electrons, a linear dc current and periodic voltages.  It should be more facilitated here by the compact form of the FR in Eq.\eqref{Sphoto} in terms of the dc noise $S_{dc}$ which has a non-trivial behavior in non-linear systems. There are in addition situations where the sources to be probed are non-periodic, such as a random non-gaussian radiation (see Eqs.\eqref{p_average},\eqref{Sphoto_cumulant}).  Without knowledge of the underlying model, one could measure the noise both in absence and in presence of the sources, then extract $\bar{P}(\omega')$ by varying the dc drive $\omega_J$. 

Indeed as $\bar{P}(\omega')=|p(\omega')|^2$ in Eq.\eqref{Sphoto} hides the phase of $p(\omega')$, it is more efficient to consider PASN at a finite frequency $\omega$, where non-diagonal terms $p(\omega')p^*(\omega'+\omega)$ enter (see Ref.\onlinecite{ines_cond_mat}). Interestingly, we have also obtained these non-diagonal terms in the differential of the PASN with respect to $\varphi(\omega)$, given by  Eq.\eqref{FDT2_periodic_zero_derivative} (or the ac voltage in case Eq.\eqref{eq:JR} holds). In order to evaluate differentials, the phase of ${p}(t)$ has be known, so that this procedure applies when one needs to determine its time dependent amplitude (e.g. for tunneling or a Josephson energy). Nonetheless, one could superimpose a controlled phase $\varphi_{a}(t)$ on an unknown phase $\varphi(t)$, then take the differential in Eq.\eqref{FDT2_periodic_zero_derivative} in the limit $\varphi_{a}(t)=0$, such that $p(\omega')$ on the r.h.s. becomes determined only by $\varphi(t)$. Notice that one can also superimpose a periodic $\varphi_a(t)$ on top of a non-periodic $\varphi(t)$.

Now one could probe directly a small enough $\varphi(t)$, using the second order expansion in Eq.\eqref{FDT2_periodic_zero_second_derivative_zero_ac}. This is especially easier when one applies a sine phase without knowing its amplitude $\varphi_{ac}$, for instance renormalized by interactions while keeping the same form: $\varphi(t)=\varphi_{ac}\cos \Omega_0t$. Then, given an arbitrary $\omega_J$, one needs to measure both the PASN and the dc noise and to consider the ratio in Eq.\eqref{expansion_sine}. 

Another spectroscopy scheme, valid in the case of a thermal distribution, could be based on exploiting the thermal contribution on the r.h.s. of Eq.\eqref{FDT2_periodic_zero_thermal_n}, $2P_{-N}k_B T G_{dc}(T)$. For each $N$ (thus a dc voltage), looking at the unique term in the noise which depends on $T<\hbar \Omega_0/k_B$ provides $P_{-N}$, provided one knows $G_{dc}(T)$. 
 
\subsection{Robust determination of the fractional charge}
\label{subsec_fractional_charge}
An important family of applications of our approach consists into robust methods we have proposed for the determination of the fractional charge in the FQHE,\cite{ines_eugene,ines_cond,ines_PRB_2019,ines_PRB_R_noise_2020} and implemented experimentally to determine $e^*=e/5$ at $\nu=2/5$ in Ref. [\onlinecite{glattli_photo_2018}] and $e^*=e/3$ at $\nu=2/3$ in Ref.[\onlinecite{ines_gwendal}]. They are more robust compared to the dc poissonian shot-noise \cite{saminad} in Eq.\eqref{poisson_T}. In particular they don't require thermalized states nor high voltages which could induce heating. They are based on looking at the noise argument rather than a proportionality factor, as the key step is to determine the Josephson frequency $\omega_J$, which yields the charge $e^*$ in case the relation in Eq.\eqref{josephson} holds. The method based on the NE FR in Eq.\eqref{Sphoto} works better if the dc noise has a singular behavior close to zero, which corresponds to a locking: $\omega_J=N\Omega_0$. Such a singularity becomes more pronounced by taking the second derivative $\delta^2S_{ph}(\omega_J)/\delta^2\omega_J$, formed by a series of peaks around $N\Omega_0$. Nonetheless, if one deals with an initial thermal $\rho_0$, a low enough temperature is required to preserve these peaks, which would be otherwise rounded by thermal effects when $|\omega_J-N\Omega_0|< k_BT/\hbar$ (see the second term on the r.h.s. of Eq.\eqref{FDT2_periodic_zero_thermal_n}).

We propose here a more direct method which does not rely on such a singular behavior nor low temperatures, and equally valid for a NE $\rho_0$. It is based on the FR for the second differential of the PASN in Eq.\eqref{FDT2_periodic_zero_second_derivative}. By comparing both sides, here determined by a unique function, the PASN, one can infer the value of $\omega_J$ that ensures the equality. This would be easier in the limit of a small cosine modulation, using Eq.\eqref{FDT2_periodic_zero_second_derivative_zero_ac}: one can plot the difference of both sides as a function of $\omega_J$ and look at the value of $\omega_J$ for which it vanishes. 

We have also derived a similar relation for the PAC. \cite{ines_eugene} Nonetheless, the PAC becomes trivial for a linear dc current (see Eq.\eqref{photocurrent_linear}), as is often the case in the experimental works aiming to determine the fractional charge \cite{glattli_photo_2018,ines_gwendal}, thus motivating their recourse to methods we proposed that are based on noise.\cite{ines_cond,ines_degiovanni_2016} Thus the measured dc current does not obey a power law behavior as predicted by the effective theories. This illustrates precisely the power of such methods, which are independent on the underlying microscopic description of the edge states, as long as it can be cast in the form of Eq.\eqref{Hamiltonian}. 


 \section{Discussion and conclusion}

We have studied the noise generated by radiation fields operating in a large family of physical systems, such as a QPC in the FQHE or the IQHE with interacting edges, as well as quantum circuits formed by a JJ, NIS or dual phase slip JJ strongly coupled to an electromagnetic environment. We have related the PASN in a universal manner to its counterpart in a dc regime characterized by a NE distribution, similarly to relations obeyed by the finite-frequency current for ac drives \cite{ines_PRB_2019} and finite-frequency noise in the dc regime \cite{ines_PRB_R_noise_2020}.
The NE FRs unify higher dimensional and one-dimensional physics, though the latter is atypical as it is drastically affected even by weak interactions. They also unify previous works based on specific models and an initial thermal equilibrium.\cite{glattli_degiovanni_PRB_2013,glattli_levitons_physica_2017,crepieux_photo,martin_sassetti_prl_2017}

We can transpose to the PASN various methods based on the PAC and addressed in Refs.\onlinecite{ines_eugene,ines_PRB_2019,ines_eugene_detection}, in particular to probe the fractional charge or to detect current cumulants of a non-gaussian source, though the implementation is not identical due to different properties of current and noise. Indeed, in case the dc current is linear and $|\bar{p}(t)|=1$, the PAC becomes trivially equal to the dc current and the PASN offers a non-trivial alternative, as is the case in two situations arising in the IQHE and FQHE.

One the one hand, interactions between edge states still play an important role in the IQHE, which is addressed in many works through the plasmon scattering approach. \cite{ines_schulz_group,ines_epj,elec_qu_optics_degiovanni_2017_cite,plasmon_ines_IQHE_HOM_feve_Nature_2015_cite,sassetti_levitons_IQHE_2020} But bosonized models justify the linearity of the dc current through a spatially local QPC, which justify the recourse to scattering approach for independent electrons in those works. Notice that all these hypothesis are not required within our approach, as we can deal with a possible non-linear dc current, which is the signature of the QPC.

On the other hand, in the FQHE, the dc measured current is quite often weakly non-linear in experiments aiming to probe fractional charge \cite{glattli_photo_2018,ines_gwendal} and statistics \cite{fractional_statistics_gwendal_science_2020} As the PAC is trivial, the FR for the PASN, already obtained in Ref.\onlinecite{ines_cond_mat}, has been fruitful for an experimental determination of the fractional charge at $\nu=2/5$ \cite{glattli_photo_2018} as well as for the analysis of two-particle collision experiments. \cite{glattli_imen_2022,Imen_thesis} 
In those experimental works where effective theories such a the TLL are not in accordance with the observed dc current, our methods have the advantage to be robust with respect to the underlying microscopic description and non-universal features, such as edge reconstruction or absence of edge equilibration. In the present paper, we have proposed a more advantageous method based on a second differential of the PASN, which would be easier to exploit in the limit of a weak sine voltage.

We have also discussed how the NE FR is potentially relevant to shot-noise spectroscopy as well as to current cumulant detection. \cite{ines_PRB_2019,ines_eugene_detection}
 We have found that the excess noise can be negative in a non-linear SIS junction. Thereby the qualification of "photo-assisted" is not universally relevant: the PASN can be reduced by an ac voltage superimposed on a dc one. This feature is at odd with a theorem by L. Levitov {\it et al} \cite{keeling_06_ivanov} which is restricted to a linear dc current. Such a theorem was at the heart of characterizing minimal excitations for an initial thermal equilibrium at low temperatures. We have provided an alternative characterisation. Showing that the PASN is super-poissonian whatever is the NE initial distribution, the lorentzian profile of the voltage is precisely the one which leads to a poissonian PASN when the system is in the ground many-body state. For hierarchical states of the FQHE, we showed that the charge carried by minimal excitations is still depending on the fractional charge, and that the PASN is superpoissonian, as will be exposed in more details in a separate publication.

Finally, compared to the dc regime, additional limitations of the approach arise. These are mainly due to possible couplings to time dependent forces or boundary conditions which have to be incorporated into a unique complex function, for instance through unitary transformations of the Hamiltonian. This needs to be checked in presence of a supplementary tunneling point between edges with a different ac voltage one could absorb through a translation of the bosonic fields, but seems more difficult to ensure for the multiple mixing points addressed in Ref.\onlinecite{glattli_imen_mixing_2022}.
Also an interferometer with multiple QPCs driven by different ac voltages is not expected to enter within our domain of validity. The present approach would still offer a test in the limiting cases of identical ac voltages, or of a dominant tunneling through one QPC. 

In the quantum Hall regime with an almost open QPC, one usually measures correlations between chiral edge currents. These are determined by the backscattering PASN owing to current conservation at zero frequency. 
But it might need to be completed  \cite{trauzettel_group_bis,lee_levitov,photo_grabert_noise}
in NE setups, where the perturbative approach can still be useful, as addressed in future works. 

Finally, an important open question consists into finding the criteria for minimal excitations which would go beyond the second-order perturbation we have carried on. 

Acknowledgments: The author thanks R. Deblock, P. Degiovanni, G. F\`eve, I. Taktak, C. D. Glattli and B. Dou\c {c}ot for discussions. She also thanks E. Sukhorukov for inspiring discussions during past collaborations. 

\appendix
\section{Derivation of the NE FR}
\label{app:noise}
This appendix provides a detailed derivation of the relation obtained in Eq.\eqref{Sphoto}. In order to express the PASN in Eq.\eqref{Sdefinition_zero} to second order of perturbation with respect to ${A}$, we don't need an expansion of the S-matrix, as ${S}$ is already of second order. Thus we can directly replace $\delta\hat I_{\mathcal{H}}(t)$ by $\hat
I_{\mathcal{H}_0}(t)$, or, in Eq.\eqref{eq:current}, 
${A}_{\mathcal{H}}(t)$ by ${A}_{\mathcal{H}_0}(t)=e^{i{\mathcal H}_0
t} {A}\,e^{-i{\mathcal H}_0 t}$. Then the effect of the ac drive factorizes: 
\begin{subequations}
\begin{align}
\label{https://www.overleaf.com/projectS_time}
 {S}(\omega_{J};t,\tau)&=e^* e^{-i\omega_J\tau}{p}\left(t+\frac{\tau}2\right){p}^*\left(t-\frac{\tau}2\right)I_{\rightarrow}(-\tau)+e^{i\omega_J\tau}{p}^*\left(t+\frac{\tau}2\right){p}\left(t-\frac{\tau}2\right)I_{\leftarrow}(\tau).
 \end{align}
 \end{subequations}
 The two correlators $I_{\rightarrow}(\tau), I_{\leftarrow}(\tau)$ determine all observables associated with the current in Eq.\eqref{eq:current} to second order perturbation; they keep track of unspecified Hamiltonian and initial stationary NE density matrix ${\rho}_0$, thus depending only on time difference $\tau$ \cite{ines_PRB_2019,ines_PRB_R_noise_2020}:
\begin{subequations}
\begin{align}
\label{Xup_Xdown}
\hbar^2 I_{\rightarrow}(\tau)&=e^*\langle{A}_{\mathcal{H}_0}^{\dagger}(\tau){A}_{\mathcal{H}_0}(0)\rangle \;\;\;\;\; \hbar^2 I_{\leftarrow}(\tau)=e^*\langle{A}_{\mathcal{H}_0}(0){A}_{\mathcal{H}_0}^{\dagger}(\tau)\rangle.
\end{align}
\end{subequations}
Notice that the Fourier transforms $I_{\rightarrow},I_{\leftarrow}(\omega)$ are real because both functions verify: $X^*(\tau)=X(-\tau)$ \cite{ines_philippe_group,ines_cond_mat,ines_PRB_2019}. Here $X(\omega)=\int d\tau e^{i\omega\tau} X(\tau)$, as the measurement time $T_0$ delimits only integration over $t$. This implements the first important step underlying the derivation of various NE perturbative relations. 

 As time-translation invariance is broken, double-Fourier transform  introduces two frequencies, $\omega,\Omega$ : 
\begin{equation}\label{Sdefinition_fourier}
{S}(\omega_{J};\omega,\Omega)=\int_{-T_0/2}^{T_0/2} \frac{dt}{T_0} \int_{-\infty}^{\infty}d\tau e^{i\Omega t} e^{i\omega \tau} {S}(\omega_{J};\tau,t).
 \end{equation} 
Focussing here on $\Omega=\omega=0$, and letting $S_{ph}(\omega_{J})=S(\omega_{J};0,0)$, we obtain the PASN in terms of the Fourier transforms of the two correlators in Eq.\eqref{Xup_Xdown}:
 \begin{subequations}
\begin{align}
\label{S_frequency}
 S_{ph}(\omega_{J})&=e^*\int  \frac{d\omega'}{\Omega_0} |p\left(\omega'\right)|^2\left[I_{\rightarrow}(\omega_J+\omega')+I_{\leftarrow}(\omega_J+\omega')\right].
  \end{align}
 \end{subequations}
We have defined: $p(\omega)=\int_{-T_0/2}^{T_0/2}e^{i\omega t} p(t)dt/T_0$. An additional term, not considered here, and due to a possible singularity $p_{dc}\delta(\omega)$ (such as is the case for a single lorenztian pulse), can be found in a similar fashion as for the PAC  in Ref.\onlinecite{ines_PRB_2019}.  

It is useful to recall how $I_{\rightarrow},I_{\leftarrow}$ determine as well the expressions of current average and zero-frequency noise in the dc regime, i.e. at ${p}(t)=1$, for which we choose the subscript $dc$ \cite{ines_cond_mat,ines_degiovanni_2016,ines_PRB_R_noise_2020}:
\begin{subequations}
\begin{align}
I_{dc}(\omega_{\mathrm{J}})&=	
I_{\rightarrow}(\omega_{\mathrm{J}})-I_{\leftarrow}(\omega_{\mathrm{J}})\,
	\label{formal_average}\\
S_{dc}(\omega_J)/e^{*} &=
I_{\rightarrow}(\omega_J)+I_{\leftarrow}(\omega_J)\,.\label{noise_DC_initial:time}
\end{align}
\end{subequations}
Thus, using as well a spectral decomposition, we can view $I_{\rightarrow}$ and $I_{\leftarrow}$ as transfer rates in opposite 
directions, whose difference yields the dc current, while their superposition evaluated at two effective dc voltages yields the FF noise. For a Josephson junction in series with an electromagnetic environment, they play the role of the $P(E)$ function for initial thermal states, offering its two counterparts for NE states. 

We stress that, contrary to the majority of previous studies on time-dependent transport, the two correlators  $I_{\rightarrow},I_{\leftarrow}$ are not necessarily linked: one can have $I_{\rightarrow}(\omega)\neq I_{\leftarrow}(-\omega)$ and they don't obey a detailed balance 
equation if we don't consider
initial thermal states. Therefore, $I_{\rightarrow},I_{\leftarrow}$ are, in full generality, two independent functions. 

Next, in order to derive the FR relating the noise under the drive $p(t)$ to that in the dc regime, $S_{dc}$, we compare the two expressions respectively given by Eq.\eqref{S_frequency} and Eq.\eqref{noise_DC_initial:time}. 

One can then see that the combination of the NE correlators in the integral of Eq.\eqref{S_frequency} is nothing but the noise in the dc regime, evaluated at an effective dc drive given by $\omega_J+\omega$. This leads to Eq.\eqref{Sphoto}.

\section{Periodic drives}
\label{app_periodic}
This appendix is devoted to detail the PASN and the PAC in presence of a periodic ${p}(t)$ at a frequency $\Omega_0$. Then the integral in Eq.\eqref{Sphoto} reduces to a sum over $\omega'=l\Omega_0$ for integer $l\geq 1$, which leads to Eq.\eqref{FDT2_periodic_zero}.  
 
This relation is similar to the PAC in Eqs.\eqref{average_current},\eqref{Iphoto}: \cite{ines_eugene,ines_PRB_2019}
\begin{equation}\label{Iphoto_periodic}
{I}_{ph}(\omega_{J})=\sum _{l=-\infty}^{+\infty}P_l   I_{dc}(\omega_{J}+l\Omega_0).\end{equation}

We notice that when $|{p}(t)|\neq 1$, we have $\sum _{l=-\infty}^{+\infty} {p}_{l}p^*_{l+k}=F.T.\left[|p(t)|^2\right]_k,$
the Fourier transform at $k\Omega_0$ of $|p(t)|^2$. In particular, for $k=0$, we have:  \begin{equation}\label{sum_Pl}\sum_{l=-\infty}^{l=+\infty} {P_l}=<|p(t)|^2>_{T_0},\end{equation} where average refers to that over a period (see Ref.\onlinecite{ines_PRB_2019} for a non-periodic ${p}(t)$). When $|{p}(t)|= 1$, we recover the orthogonality: 
\begin{equation}|{p}(t)|= 1\implies\sum _{l=-\infty}^{+\infty} {p}_{l}p^*_{l+k}=\delta_k,\label{orthogonality}\end{equation}
where $\delta_k$ is the Kronecker sign.

\section{Initial thermal equilibrium distribution}\label{sec_thermal}
 This appendix gives a more detailed expression of the PASN for an initial thermal distribution. By injecting the expression of the dc noise in Eq.\eqref{poisson_T} it into Eq.\eqref{Sphoto}, the PASN becomes totally determined by the NE dc current: 
\begin{equation}
S_{ph}(\omega_J)=e^*\int \frac{d\omega'}{\Omega_0}\bar{P}(\omega'-\omega_J)  \coth\!\left(\frac {\hbar\omega'} {2k_BT}\right) I_{dc}(\omega').\label{Sphoto_thermal}
\end{equation}
Our expression is different from the one derived by L. Levitov {\it et al} for a QPC in the FQHE \cite{keeling_06}, and which we recover only for a linear dc current : $I_{dc}(\omega')=\hbar G_{dc}\omega'/e^*$. 
For a periodic drive, we get:
\begin{equation}\label{FDT2_periodic_zero_thermal}
{S}_{ph}(\omega_{J})= e^*\sum _{l=-\infty}^{l=+\infty} P_l \;\coth\!\left[\frac {\hbar(\omega_J+l\Omega_0)} {2k_BT}\right] I_{dc}(\omega_{J}+l\Omega_0).\end{equation}

Let us now specify further to the case $\omega_J=N\Omega_0$. Since one deals here with an equilibrium thermal distribution, one has  $\omega_J I_{dc}(\omega_J)\geq 0$ even when $I_{dc}$ is not odd (see Ref. \onlinecite{ines_PRB_2019}).
If we consider now the NE quantum regime at $\Omega_0\gg k_B T/\hbar$, we obtain Eq.\eqref{FDT2_periodic_zero_thermal_n} in the text. Though we have a NE PASN, an equilibrium thermal noise weighted by $P_{-N}$ arises, given by $P_{-N}S_{dc}(0)$. It is due to a vanishing effective dc frequency when the many-body state at energy $N\Omega_0$ emits $N$ photons. For independent electrons, it was interpreted 
as a reduced thermal contribution from the reservoirs.\cite{glattli_levitons_nature_13}  In case the dc current is linear, one has generically $G_{dc}(T)=G_{dc}$ constant, so that the contribution of $l=-N\pm 1$, proportional to $G_{dc}\hbar\Omega_0$, dominates $k_BT G_{dc}$. Nonetheless, one needs also to compare $P_{-N\pm 1}$ to $P_{-N}$, so we cannot ignore the term $P_{-N} k_BT G_{dc}$ independently on the ac voltage.    

One cannot ignore it as well in non-linear conductors, where $G_{dc}(T)$ depends on $T$, as is the case of effective theories for the FQHE, addressed in a separate paper. 

\section{Three choices for the excess PASN}

Here we discuss the choice of the excess PASN. 
\label{app_excess}
This appendix aims to discuss the excess noise. 
One often deals with excess noise in order to get rid of undesirable noise. In the dc regime, it is often defined as:
\begin{eqnarray}\Delta S_{dc}(\omega_J)&=&S_{dc}(\omega_J)-S_{dc}(0)\label{def_excess_dc} 
\end{eqnarray} Recall that in a NE setup with couplings to dc voltages independent from $\omega_J$, $S_{dc}(0)$ is finite even when all temperatures are set to zero, and is therefore different from the thermal equilibrium noise.

For the PASN, there isn't a single convention as it depends on which reference is chosen in a given experimental context.
In Eq.\eqref{second_choice}, we have chosen as a reference the dc noise in presence of the same dc voltage. 

One could also adopt a second choice, by substracting the same reference as that in Eq.\eqref{def_excess_dc}, \begin{eqnarray}\Delta {S}_{ph}(\omega_J)&=&{S}_{ph}(\omega_J)-S_{dc}(0)\label{def_excess_noise_ph}
\end{eqnarray}
 This yields, focussing on a periodic drive at a frequency $\Omega_0$ (see Eq.\eqref{FDT2_periodic_zero}):
\begin{equation}\label{FDT2_periodic_zero_excess}
\Delta{S}_{ph}(\omega_{J})=\sum _{l=-\infty}^{+\infty} P_l   \Delta S_{dc}(\omega_{J}+l\Omega_0)+\left(\sum _{l=-\infty}^{+\infty} P_l -1\right )S_{dc}(0).\end{equation} 
The relations between the two choices is given by :
\begin{equation}\label{app_second_choice}
\underline{\Delta} S_{ph}(\omega_J)=S_{ph}(\omega_J)-S_{dc}(\omega_J)=\Delta S_{ph}(\omega_J)-\Delta S_{dc}(\omega_J).
\end{equation}

Excess noise is expected to have a positive sign, as noise should increase with additional voltage sources. This is indeed not systematic in the dc regime, as we have shown for zero or finite frequency noise for non-linear conductors.\cite{dolcini_07,ines_philippe_group} In the text, \ref{subsec_SIS}, we showed that the choice in Eq.\eqref{second_choice} can lead to a negative sign in a SIS junction. In a separate work, we will show that Eq.\eqref{FDT2_periodic_zero_excess} has a negative sign in the FQHE.

In view of the super-poissonian noise in Eq.\eqref{super_poisson}, a third choice guarantee a positive sign : $S_{ph}(\omega_J)-e^*|I_{ph}(\omega_J)|$. 
But such a choice is not so advantageous. 
If the dc current is nonlinear, one would need to measure the non-trivial PAC. In addition, substracting a noise reference is more convenient to get rid of undesirable sources which affect the PASN in a different manner from the PAC . For instance, if one takes the zero dc voltage limit, one has $I_{ph}(0)=0$ in case $I_{dc}$ is odd and $P(\omega')=P(-\omega')$, but has still a finite $S_{ph}(0)$. \\
A similar choice was given in Ref.\onlinecite{martin_sassetti_prl_2017}. Restricted to a thermal equilibrium and to the TLL model, that work recovered the super-poissonian PASN of Ref.\cite{ines_cond_mat}. This motivated the authors to define the excess noise as: ${S}_{ph}(\omega_{J})-e^*\coth\left[\beta \omega_{J}/2\right]I_{ph}(\omega_J)$, whose sign is however not well determined. In case the dc current is linear (see Eq.\eqref{photocurrent_linear}), this amounts to adopt the second choice, Eq.\eqref{second_choice}. Such a definition was intended to cancel thermal contributions, but indeed cancels only the contribution of $l=0$ in Eq.\eqref{FDT2_periodic_zero_thermal_n}, and not the term $P_{-N}S_{dc}(0)$. 
 

\begin{thebibliography}{10}
\bibitem{note_voltages}
One can define $V_{dc},V_{ac}$ as effective dc and ac voltages to which we
  might refer, though different from the applied ones. Then the choice of $e^*$
  would depend on the context.
\bibitem{note_scaling}
The scaling dimension is defined through the Green's function for the
  associated quasiparticle field $\Psi$:
  $\left<\Psi^{\dagger}(x,t)\Psi(x,0)\right>\simeq (\omega_F t)^{-\delta}$,
  where $\hbar\omega_F$ is an energy cutoff. The bahavior of the correlator
  associated with $A$ is then $(\omega_J/\omega_F)^{2\delta-1}$, thus is
  greater when $\delta$ is smaller.
\end{thebibliography}

\end{document}